\documentstyle[prd,aps,multicol,epsfig]{revtex}

\newcommand{\be}{\begin{equation}}    
\newcommand{\ee}{\end{equation}}
\newcommand{\beq}{\begin{eqnarray}}
\newcommand{\eeq}{\end{eqnarray}}

\def\op{ \ $ }
\def\cl{$ \ }

\def\nn{\nonumber}

\newcommand{\df}{\delta F}
\newcommand{\q}{Q_{e}}
\newcommand{\dr}{\delta R}

\newcommand{\rt}{\tilde{r}}

\newcommand{\dps}{\delta\psi}
\newcommand{\dmt}[1]{\delta\mu_{3 #1}}
\newcommand{\dmd}[1]{\delta\mu_{2 #1}}
\newcommand{\dnu}{\delta\nu}
\newcommand{\enu}{e^{\nu}}
\newcommand{\enud}{e^{2\nu}}

\newcommand{\rtocc}{(\log \rt e^{-\nu})_{,r}}
\newcommand{\rtoc}{(\log \rt)_{,r}}
\newcommand{\rtopc}{(\log \rt e^{\nu})_{,r}}
\newcommand{\rtocco}{(\log \rt e^{2\nu})_{,r}}
\newcommand{\ze}[1]{#1^{(0)}}

\begin{document}

\draft

\title{Quasi-normal modes of charged, dilaton black holes}

\author{Valeria Ferrari$^{(1)}$\thanks{Email address: Valeria.Ferrari@roma1.infn.it},
Massimo Pauri$^{(2)}$\thanks{Email address: pauri@pr.infn.it} and 
Federico Piazza$^{(3)}$\thanks{Email address: piazza@mi.infn.it}}
\address{$^{(1)}$Dipartimento di Fisica, Universit\'a di Roma, ``La Sapienza" and \\
I.N.F.N. Sezione di Roma1, Piazzale Aldo  Moro 5, Roma, Italy\\
$^{(2)}$Dipartimento di Fisica, Universit\'a di Parma and \\
I.N.F.N. Gruppo Collegato di Parma, Viale delle Scienze, 43100, Parma, Italy\\
$^{(3)}$Dipartimento di Fisica, 
Universit\'a di Milano and
I.N.F.N. Sezione di Milano, Via Celoria 16, 20133 Milano, Italy\\
}
\date{Received 30 May 2000}
\maketitle

\begin{abstract} 

In this paper we study the perturbations of the charged, dilaton 
black hole, described by the solution of the low
energy limit of the  superstring action found  by
Garfinkle, Horowitz and Strominger.
We compute the complex frequencies of the quasi-normal modes of this black
hole, and  compare the results with those obtained for a
Reissner-Nordstr\"{o}m and a Schwarzschild black hole. 
The most remarkable feature which emerges from this study is 
that the presence of the dilaton breaks the \emph{isospectrality} of 
axial and polar perturbations, which characterizes both Schwarzschild
and Reissner-Nordstr\"{o}m  black holes.
 
\end{abstract}

\section{INTRODUCTION} \label{intro}

The  study of the effects produced by the coupling between 
gravity and scalar fields is a rather hot issue, since 
a definite prediction of superstring theory is just the existence
of a scalar field, namely the \emph{dilaton}. 
For instance, this coupling  could have played an important role 
during the early phases of the life of the Universe.
Indeed, it has been shown  that a possible cosmological implication
of the dilaton is the so-called ``pre-big-bang'' scenario\ \cite{PBB},
characterized by a dilaton-driven inflationary period, that should have
left its peculiar fingerprints on the cosmic gravitational wave 
background\ \cite{maggiore}.
On the experimental side, recent investigations on the interaction 
of scalar waves with gravitational detectors have shown that a 
scalar component of the gravitational radiation should excite the 
monopole mode of a 
resonant spherical antenna\ \cite{maura} and also give rise to 
specific correlations between the signals 
revealed by a resonant sphere and an interferometer\ \cite{nicolis}.

Among all possible astrophysical sources of gravitational waves,  black holes 
should have the most typical and in principle recognizable frequency 
spectrum. According to the Einstein theory of gravity,
a black hole which dynamically interacts  with its surroundings 
emits wave bursts having shape and intensity which initially depend
on the features of the external perturbations.
As a late-time effect, however, it generates  gravitational wave 
trains having characteristic frequencies which are 
\emph{independent} of the initial perturbations,
the so-called \emph{quasi normal modes}\ \cite{qnm}. 
And we know from the theory that these frequencies can only depend  
on the three parameters characterizing a black hole, 
namely the mass, the charge and the angular momentum.

The question whether a black hole can support an additional scalar 
degree of freedom 
and still generate a regular,  asymptotically flat,
spacetime free of naked singularities or any other
kind of pathological behaviour has been widely investigated in the
literature. Having in mind the ``no-hair theorem'' program, 
Bekenstein\ \cite{bek1} in 1972 showed that  black holes cannot
support a \emph{minimally coupled} scalar field,
if the solution is required to be asymptotically flat.
Subsequently, he developed  a procedure to generate black hole solutions 
surrounded by  a \emph{conformal} scalar field\ \cite{bek3},
but they proved  to be \emph{unstable} against radial perturbations\ \cite{bron}. 
In 1990 Ferrari and Xanthopoulos \ \cite{valexan}, 
using a Kaluza-Klein approach, derived  equations 
describing a gravitational field coupled with a massive scalar field,
obtained the conformal structure of the metric, and discussed several kind
of possible couplings. Yet, no definite answer was given to the
question whether a satisfactory black hole can exist surrounded 
by a massive scalar field.

The search for regular exact solutions describing black holes endowed
with a 
scalar field got a formidable boost  when it became clear that, 
as long as the curvature is small, all vacuum solutions of general 
relativity are also approximate solutions of string theory.
As a matter of fact, charged
black hole solutions  coupled to a scalar field, viz a {\it dilaton}
 in this context, can be obtained  by the action (we use geometric units throughout)
\be\label{theaction}
S = \int d^{4}x \sqrt{-g}\ [R - 2(\nabla \phi )^{2} + e^{-2a\phi} F^{2}],
\ee
which, by variations, gives the equations:
\begin{equation}\label{maxxeq}
\nabla_{\mu}( e^{-2a\phi} F^{\mu\nu})=0,
\end{equation}
\begin{equation}\label{dilaeq}
\nabla^{2} \phi - \frac{1}{2} e^{-2 a \phi} F^{2}=0,
\end{equation}
\begin{equation}\label{einseq}
R_{\mu\nu} =2\nabla_{\mu}\phi\nabla_{\nu}\phi
 - 2e^{-2a\phi} F_{\mu\rho}F_{\nu}^{\
\rho} + \frac{1}{2} g_{\mu\nu} e^{-2a\phi} F^{2},
\end{equation}
where $F^{2} = F_{\mu\nu}F^{\mu\nu}$ is the first Maxwell 
invariant and $a$ is a non-negative real constant 
regulating the strength of the 
coupling between the dilaton and the Maxwell field.
This theory has a number of interesting limiting cases.
The limit $a=0$ corresponds 
to the ordinary Einstein-Maxwell theory plus a Klein-Gordon
scalar field with zero mass. 
The case $a=\sqrt{3}$ is the four-dimensional reduction of Kaluza-Klein
theory. Finally, for $a=1$ the action (\ref{theaction})
describes the tree-level low energy limit of superstring theory in the so-called
\emph{Einstein frame}. A large class of black hole solutions of this theory 
in an arbitrary 
number of dimensions has been found by Gibbons and Maeda\ \cite{gibb1}. 
Their results were next specialized to four dimensions 
by Garfinkle, Horowitz and Strominger\ \cite{ghs} (GHS)
who studied a solution describing a  spherically symmetric, 
charged, dilaton black hole.
In the GHS solution  the
electric charge and the dilaton are not independent parameters: when the 
charge is set equal to zero the dilaton also disappears, 
and the solution reduces
to  that of a Schwarzschild  black hole. In a sense, this is a consequence 
of the ``no-hair theorems'' which limits  the number of free parameters 
of a black hole to three.

The equations satisfied by small perturbations of this solution
have been derived and studied by Holzhey and Wilczek\ \cite{HW} in the line of 
the general treatement given by Chandrasekar\ \cite{MT}. They first 
showed that it is possible to reduce the perturbed
equations to five decoupled  wave equations with potential barriers, and,
as a by-product of their analysis, they argued that the  
GHS solution is stable under external small perturbations.

In this paper we take a further step in the study
the properties of the coupled emission of electromagnetic, 
scalar and gravitational radiation by such black holes,
computing the \emph{quasi-normal mode frequencies} of the  GHS
solution.
Comparing the spectrum of the latter with those of  
Schwarzschild and 
Reissner-Nordstr\"om,  we show that the presence of the
scalar field breaks the relevant feature of the \emph{isospectrality} of the
axial and polar perturbations.

The paper is organized as follows. In Section 2 we summarize the features 
of the GHS solution. In Sections 3 and 4 
the equations 
governing the axial and polar perturbations are discussed separately.
We shall not derive the axial equations, since there is nothing to add to
the derivation made in   Ref. \ \cite{HW}.
On the other hand, the polar equations will be discussed in greater detail.
Some misprints appearing in Ref. \ \cite{HW} are corrected and the explicit 
expression of the 
matrix whose eigenvalues are the potentials governing polar perturbations
is given in the Appendix. In Section 5 the quasi-normal frequencies 
of the GHS black hole obtained with an extended WKB approch\ \cite{iyerwill}  
are calculated for 
different values of the parameters.
Finally, in Section 6 the results obtained are discussed  and the 
main {qualitative} differences between the spectra of the dilaton
and  the Reissner Nordstr\"{o}m black holes are pointed out.

\section{THE EXACT SOLUTION}\label{sec:due}
The exact solution of equations (\ref{maxxeq})-(\ref{einseq}),
with $a=1$,  describing the 
charged dilaton black hole 
we study in this paper is\ \cite{ghs}
\be
\label{metric}
ds^{2}=\left( 1-\frac{2M}{r}\right) dt^{2} -\left(
1-\frac{2M}{r}\right)^{-1} dr
^{2}-
r\left(r-\frac{\q^{2}}{M}\right)
\left[d\theta^{2}+sin^{2}\theta d\varphi^{2}\right]
\ee
where  $M$\ and $\q$\ are, respectively, the black hole mass and  
electric charge.
$F_{t r}= \q/r^{2}$
is the only non-vanishing component of the electromagnetic tensor, and
the scalar field is related to the electric charge by the following
equation
\be
e^{2\phi}=\left(1-\frac{\q^{2}}{Mr}\right),
\label{phi}
\ee
which shows that \op \phi\cl vanishes at radial infinity.
Since the dilaton and the electric charge  are coupled through Eq. (\ref{phi}),
the Reissner N\"{o}rdstrom solution cannot be obtained as a limiting case
of the metric (\ref{metric}). On the other hand,
the  Schwarzschild solution
is recovered from Eq. (\ref{metric}) by setting $\q=0$.

It should be noted that the usual relation between radius and area of the 
spheres $t=const$, $r = const$ is obtained in terms of the modified radial variable

$$
\rt~ =~\sqrt{r\left(r-\frac{\q^2}{M}\right)},
$$
and not of $r$, by the usual relation
\op
A(r)\, = \, 4 \pi \rt^2.
\cl

The metric (\ref{metric}) appears to be singular in
$r=0$, $r =\q^{2}/M$\ and $r=2M$. The surface  $r=2M$\ is an  event
horizon, whereas on
$r=\q^{2}/M$   the curvature scalar
$$
R\ =\ 2g^{rr}(\phi_{,r})^{2}\ =\
\frac{1}{2}\left(1-\frac{2M}{r}\right)\frac{\q
^{2}}
{r(Mr-\q^{2})},
$$
diverges, showing that there is a curvature singularity.
In $r=\q^{2}/M$, the  radial coordinate  $\rt$ vanishes 
and loses its meaning for $r< \q^{2 }/M$.
Thus, a physical observer who crosses the event horizon 
terminates its jurney on the curvature singularity. 

In  order to avoid naked singularities,
in what follows we shall assume that $$2M^{2}>\q^{2}.$$

As shown in Ref.  \cite{MT}  (to be referred to hereafter as MT),
the study of the perturbations of the metric  (\ref{metric}), 
can be  restricted, without loss of generality,
to axisymmetric perturbations only. 
The appropriate metric in this case is
\be
ds^{2}= e^{2\nu}(dt)^{2} -
e^{2\psi}(d\phi-q_{2}dx^{2}-q_{3}dx^{3}-\omega dt)^{2}-e^{2\mu_{2}}(dx^{2})^{2}
-e^{2\mu_{3}}(dx^{3})^{2},
\label{metpert}
\ee
where, in the unperturbed state,
\beq
e^{2\nu} =& e^{-2\mu_{2}}&\, =\left(1-\frac{2M}{r}\right),\hspace{6mm}
e^{2\psi}=\rt^{2}sin^{2}\theta \nn \\
&e^{2\mu_{3}} &= \rt^{2},\hspace{8mm}
\omega=q_{2}=q_{3}=0\ ,
\eeq
and $(0,1,2,3)$ stand for $(t,\varphi,r,\vartheta)$.
We shall assume that, as a consequence of a generic perturbation,
the metric functions, the electromagnetic quantities and the scalar field
will experience small changes with respect to their
unperturbed values
\beq
\nu  \longrightarrow \nu+\delta\nu,\ \ 
&\mu_{2}\longrightarrow \mu_{2}+\delta\mu_{2},\ \ 
\nn
&\psi  \longrightarrow \psi+\delta\psi,\\
\mu_{3}\longrightarrow \mu_{3}+\delta\mu_{3},\ \ 
&\omega\longrightarrow\delta\omega,\ \ \nn
& q_{2}\longrightarrow\delta q_{2},\\
q_{3}\longrightarrow\delta q_{3},\ \ 
\nn
& F_{\mu \nu}\longrightarrow F_{\mu \nu}+\delta F_{\mu \nu},\ \ 
&\phi \longrightarrow \phi +\delta \phi.
\eeq
Since the perturbation is assumed to be axisymmetric, 
all perturbed quantities depend on $t$, $r$ and $\theta$ only.
As in MT, it is convenient to project Einstein's and  Maxwell's
equations onto an orthonormal tetrad frame, and assume that
all perturbed functions 
have the time dependence $e^{i\sigma t}$. The separation 
of variables is accomplished by expanding all perturbed tensors in tensorial
spherical harmonics.  These harmonics belong to two different
classes, depending on the way they behave under the angular
transformation
\op \theta\rightarrow\pi-\theta\cl and
\op \phi\rightarrow\pi+\phi.\cl
Those that transform like \op (-1)^{(\ell+1)}\cl are
termed {\it axial}, those that transform like
\op (-1)^{(\ell)}\cl are termed {\it polar}.
The perturbed equations split into two decoupled sets
corresponding to a different parity.

\section{THE AXIAL EQUATIONS}\label{sec:3}

The axial equations for \op\ell\geq 2\cl 
are obtained by perturbing the \{12\} and
\{13\}-components of the Einstein equations. The separation of variables,
can be accomplished by putting
\[
\rt^{2}e^{2\nu}\sin^{3}\theta
\left[q_{2,3}(t,r,\vartheta)-q_{3,2}(t,r,\vartheta)\right]
=Q_\ell(r,\sigma) C^{-3/2}_{\ell+2}(\vartheta) e^{i\sigma t},\]
and
\[
F_{01}(t,r,\vartheta)\sin\vartheta=
3B_\ell(r,\sigma)C^{-1/2}_{\ell+1}(\vartheta) e^{i\sigma t},
\]
where \op C^{-1/2}_{l+1}(\vartheta)\cl are the Gegenbauer polynomials.
As shown in   Ref. \ \cite{HW},
the  axial equations can be cast in the following form
\be
\label{diagcon}
\left( \frac{d^{2}}{dr_{*}^{\ 2}}+\sigma^{2} \right)\left(\begin{array}{c}
                        H_{1\ell} \\  
                        H_{2\ell} \\
                \end{array} \right)\  =\ {\bf B}\ \left(\begin{array}{c}
                        H_{1\ell} \\  
                        H_{2\ell} \\
                \end{array} \right) 
\ee
where 
\begin{equation}
{\bf B}\ =\ \frac{e^{2\nu}}{\rt^{2}}
\left[
\left(\mu^{2}+2+\frac{\q^{2}}{r^{2}}+
\frac{3\q^{4}}{4M^{2}\rt^{2}}e^{2\nu}\right)
\left(\begin{array}{cc}
                        1&0 \\
                        0&1\\  \end{array} \right)
                         +\frac{1}{r}\left(\begin{array}{cc}
                        \q^{2}/M & 2\mu \q  \\
                        2\mu \q & -6M \\ \end{array} \right)
\right],
\end{equation}
$r_{*}\cl is the tortoise coordinate defined by the equation
\be
\label{rt}
dr_{*}=e^{-2\nu}dr, 
\ee
and $\mu^{2}=(\ell-1)(\ell+2)$.
The radial functions \op H_1\cl and \op H_2\cl
(from now on we omit the index $\ell$) are related to the
perturbed metric and electromagnetic functions by the following relations
\be
Q(r,\sigma)=\rt H_{2}(r,\sigma) \hspace{16mm} 
r e^{\nu}B(r,\sigma)=\frac{H_{1}(r,\sigma)}{2\mu}\ \ .
\ee
The right hand side of Eq. (\ref{diagcon}) can be obviously diagonalized by a linear
r-independent transformation, in the form
\beq
\label{transf}
&Z_{1}^{a}=&\,{\cal L}_{1} H_{1} + {\cal L}_{2} H_{2},\\
\nn
&Z_{2}^{a}=&\,{\cal L}_{2} H_{1} - {\cal L}_{1} H_{2},
\eeq
where
\vspace{3mm}
\be
\label{tra}
{\cal L}_{1}=
\frac{\q^{2}}{2M}+3M+\sqrt{\frac{\q^{4}}{4M^{2}}+9M^{2}+\q^{2}(3+4\mu^{2})}
\quad\hbox{and}\quad
{\cal L}_{2}=2\mu\q.
\ee
Then the system (\ref{diagcon}) decouples in two wave equations: 
\be
\label{eqfin1}
\left( \frac{d^{2}}{dr_{*}^{\ 2}}+\sigma^{2} \right) Z_{i}^{a}=V_{i}^{a}Z_{i}^{a} \hspace{10mm} (i=1,\ 2),
\vspace{3mm}
\ee
where the explicit form of the effective potentials is
\vspace{3mm}
\be 
V_{1,2}^{a}=\frac{e^{2\nu}}{r\rt^{2}}\left[(\mu^{2}+2)r+\frac{\q^{2}}{r} +\frac{3\q^{4}r}
{4M^{2}\rt^{2}}e^{2\nu}+\frac{\q^{2}}{2M}-3M\pm
\sqrt{\frac{\q^{4}}{4M^{2}}+9M^{2}+\q^{2}(3+4\mu^{2})}\ \right].
\ee

Equation (\ref{tra})  shows that when $\q $ vanishes,
$ {\cal L}_{2}=0$; in this case $Z_{2}^{a}$ reduces to 
the gravitational
perturbation and  $V_{2}^{a}$
to the Regge-Wheeler potential,
whereas  \op Z_{1}^{a}\cl reduces to the pure electromagnetic
perturbation of a Schwarzschild black hole, which is known to be
independent of gravitational contributions. 
If the electromagnetic charge does not
vanish, the gravitational and electromagnetic perturbations are coupled;
a gravitational wave incident on the potential barriers induces 
the emission of electromagnetic radiation and viceversa.
On the other hand, the dilaton is not dynamically coupled with the axial perturbations.
Its effect is that of  shaping the 
effective potentials together with the electric field. In a similar manner, 
the energy density and the pressure of the matter composing a perturbed star 
determine
the potential barrier of the axial perturbations without being dynamically
coupled to the perturbed gravitational field.

The\op\ell=2\cl potentials $V_{1}^{a}$ and $V_{2}^{a}$  are plotted 
in Fig. \ref{fig:1} versus the rescaled tortoise coordinate\op r_*/M\cl 
for  different values
of the electric charge. We see that they always tend to zero at radial infinity
and at the black hole horizon, except when the charge assumes its extremal
value \op \q^2=2 M^2.\cl In this case  they take the form of a step,
reflecting all waves whose square frequency is lower than their limiting
value on the horizon. 

We also see that for non extremal black holes the maximum of the
barrier  moves towards the horizon as the electric charge increases.

\begin{figure}
\vspace{1.8cm}
\centerline{\mbox{
\psfig{figure=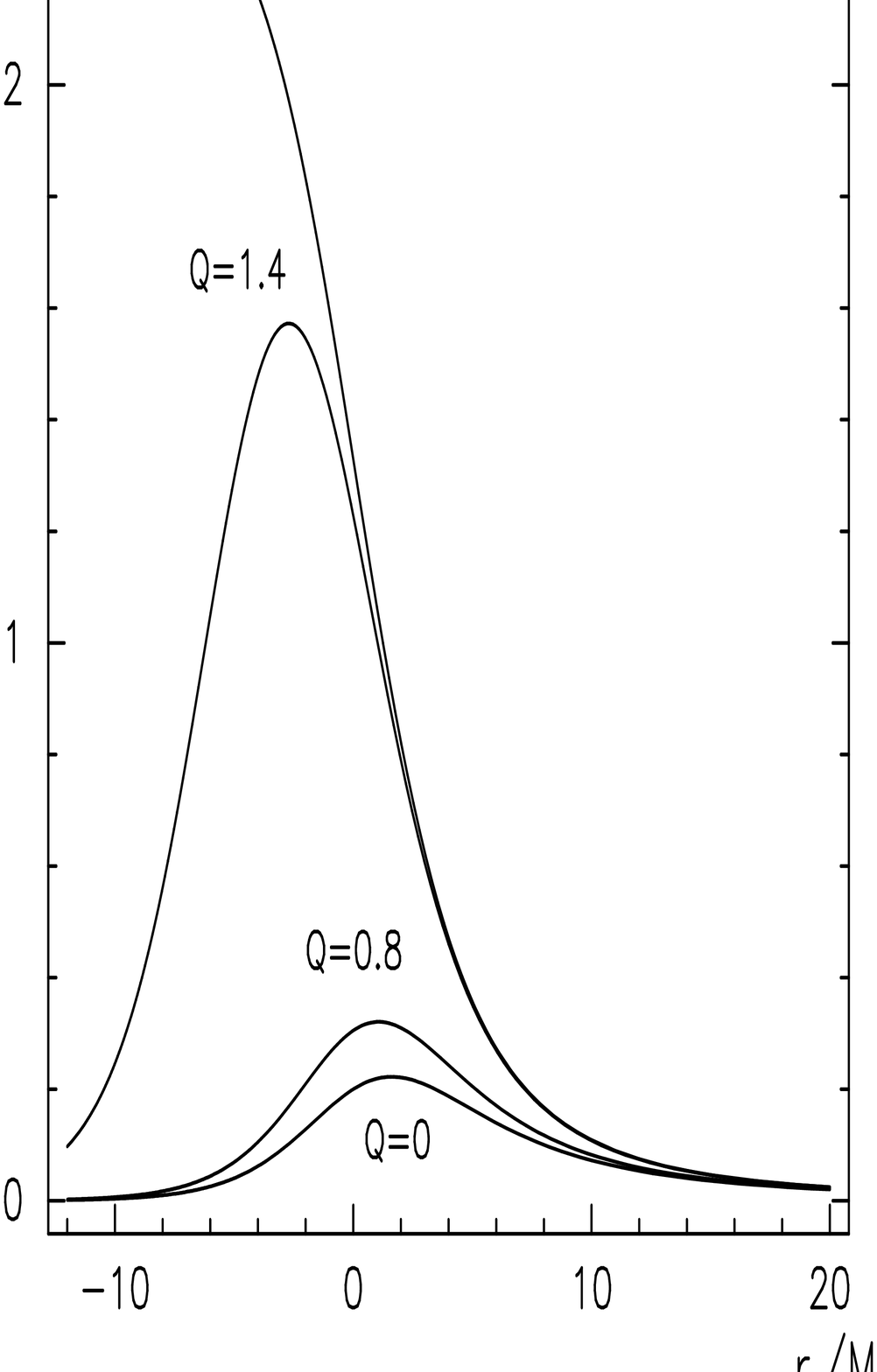,width=9cm,height=6.2cm}
\psfig{figure=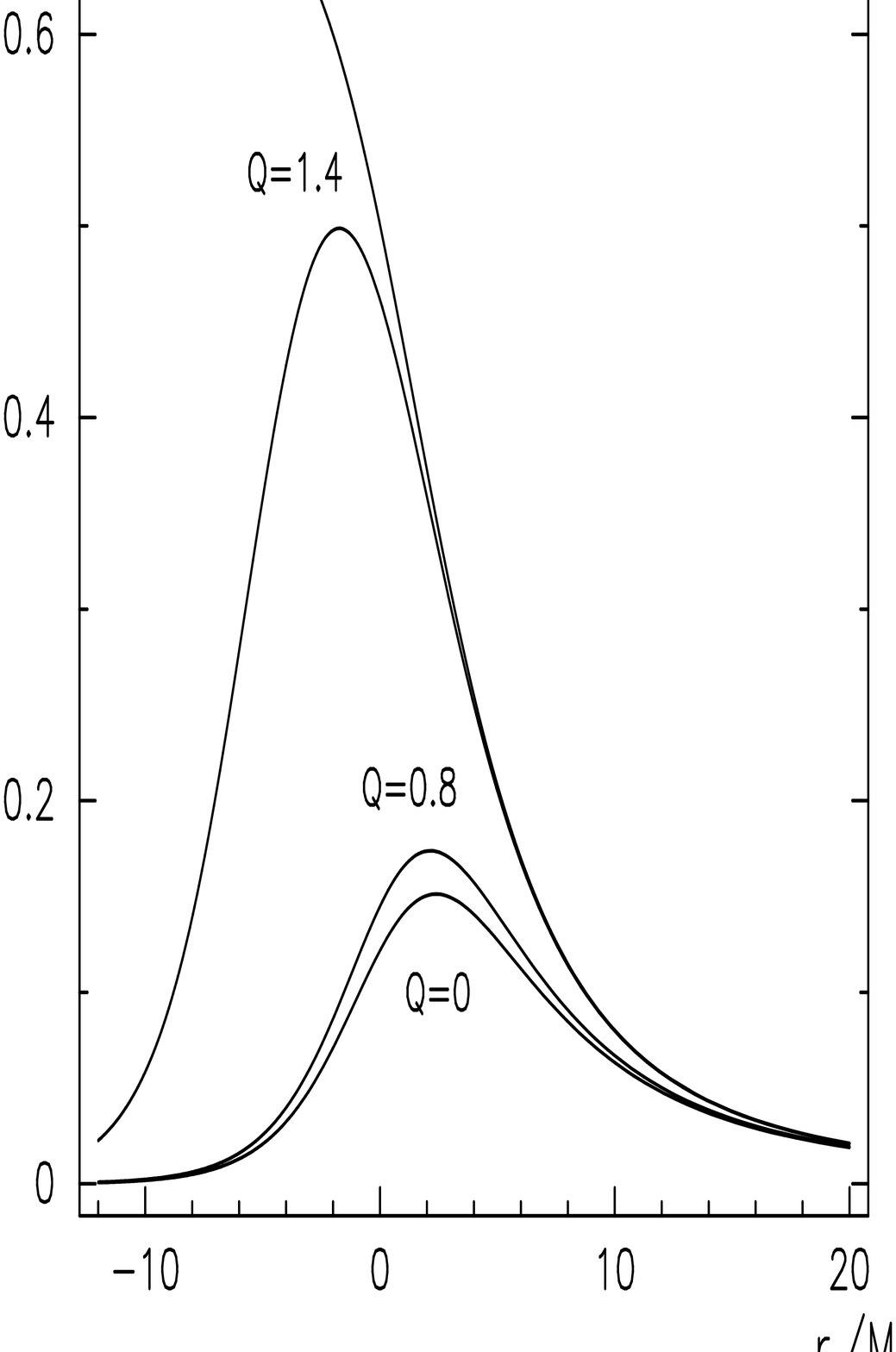,width=9cm,height=6.2cm}
}}\vspace{.5cm}
\caption{
The $\ell=2$ axial potentials $V_1^a$ (left) and $V_2^a$ (right)
are  plotted versus the rescaled tortoise coordinate for different values of
\ $ Q=Q_e/M.$\ When the charge is set equal to zero, $V_1^a$ reduces to that
of the wave equation governing the pure electromagnetic perturbations of 
a Schwarzshild black hole, whereas $V_2^a$ reduces to the Regge-Wheeler
potential.} 

\label{fig1}
\end{figure}

\section{THE POLAR EQUATIONS}\label{sec:4}

The separation of variables for polar equations is accomplished by 
requiring that the perturbed functions have the  angular
dependence  deriving from the expansion in tensor spherical
harmonics:

\beq
\label{sostituzioni1}
\dnu&=& N_\ell(r)P_{\ell}(\theta) e^{i\sigma t},\nn \\ 
\dmd{}& =& L_\ell(r)P_{\ell}(\theta) e^{i\sigma t},\nn \\ 
\dps& =&\left[ T_\ell(r)P_{\ell}(\theta) + 
2 \mu^{-2}X_\ell(r)P_{\ell}(\theta)_{,\theta}\cot\theta\right] e^{i\sigma t},\nn \\ 
\dmt{}& =&\left[ T_\ell(r)P_{\ell}(\theta) + \nn
2 \mu^{-2}X_\ell(r)P_{\ell}(\theta)_{,\theta,\theta}\right] e^{i\sigma t},\\
\df_{02}& =& -\frac{\enud \rt^{2}}{2 \q} \nn
B_{02\ell}(r)P_{\ell}(\theta) e^{i\sigma t}\nn \\
\df_{03}& =& \frac{\rt}{2 \q} 
B_{03\ell}(r) P_{\ell}(\theta)_{,\theta} e^{i\sigma t},\nn \\ 
\df_{23}& =& \frac{i e^{-\nu} \sigma \rt}{2 \q}
B_{23\ell}(r) P_{\ell}(\theta)_{,\theta} e^{i\sigma t},\nn \\ 
\delta\phi& =& \Phi P_{\ell}(\theta) e^{i\sigma t},\nn
\eeq
where $P_{\ell}(\theta)$ are the Legendre polinomials.
In the following, we shall omit the index $\ell$ in all 
radial functions.
Among the eight radial functions, 
$N$, $L$, $T$, $X$, $B_{02}$, $B_{03}$, $B_{23}$ and $\Phi$, 
the function $T$ can be eliminated from the perturbed equations by
making use of the  relation that follows from the 
$\{03\}$-component of the  perturbed Einstein equations 
\be
B_{23}-T-L+2\mu^{-2}X=0.
\ee 
From the polar components of the  Maxwell's equations (see MT, Chap. 5 Eqs.  
(165)-(167)
for the  Reissner N\"{o}rdstrom case), it is easy to  derive a 
second order equation for $B_{23}$:
\beq\label{secforb}
\left[\frac{\rt^2}{r^2}\enud(r^{2} B_{23})_{,r}\right]_{,r} +\left[\sigma^{2}\rt^{2} e^{-2\nu}-
(\mu^{2}+2)\right]B_{23}&&\nn\\
+ \frac{2 \q^2}{r^2}(2B_{23}-3L-2X-N-2\Phi)&=&0.
\eeq
The Einstein equations for 
$\dr_{02}$, $\dr_{23}$, $\delta G_{22}$ and $\dr_{11}$ give:
\beq
\left(\frac{d}{dr} + \rtocc \right)(B_{23}-L-X)-\rtoc L +\phi_{,r}\Phi& =& 0,\label{tilo1}\\
(N-L)_{,r}-\rtopc L -\rtocc N -\frac{2}{r}B_{23} + 2 \phi_{,r}\Phi& =& 0,\label{tilo2}\\
X_{,r,r}+2\rtopc X_{,r} +\frac{\mu^{2}e^{-2\nu}}{2 \rt^2}(N+L)+\sigma^{2} e^{-4\nu}X&=&0,
\label{secforx}\\
2\rtoc N_{,r}+2\rtopc (B_{23}-L-X)_{,r}-\frac{\mu^{2}e^{-2\nu}}{\rt^2}T-B_{02} &&\nonumber \\ 
-\frac{(\mu^{2}+2)e^{-2\nu}}{\rt^2}N -2\rtoc\rtocco L +2(\phi_{,r})^{2} L &&\nonumber \\
+ 2\sigma^{2}e^{-4\nu}(B_{23}-L-X)-\frac{2 \q^{2} e^{-2\nu}}{\rt^{2} r^2} \Phi - 
2\phi_{,r}\Phi_{,r}&=&0.\label{tilo3}
\eeq
Finally, the perturbed equation for the dilaton is obtained by perturbing
eq.  (\ref{dilaeq}):

\beq
\frac{1}{\rt^2}(\rt^{2}\enud\Phi_{,r})_{,r}+(\sigma^{2} e^{-2\nu} -\frac{(\mu^{2}+2)}{\rt^2}+
\frac{2 \q^{2}}{\rt^{2} r^2})\Phi \nn &&\\ 
+\enud (N-3L-2X+2B_{23})_{,r}\phi_{,r} - 2\frac{(\rt^{2}\enud\phi_{,r})_{,r}}{\rt^2}L+
\enud B_{02}&=&0.
\label{dilasos}
\eeq
The seventh order  linear system composed by equations 
(\ref{secforb})-(\ref{dilasos}) has been shown to be reducible
to three  Schr\"{o}dinger-like
equations by using the following procedure.
It is well-known that the order of a system of linear 
differential equations  can be reduced, whenever a particular solution
of that system is known.  
A general algorithm for deriving the particular solution which makes the reduction 
possible as been obtained by Xanthopulos \ \cite{xan}.
Holzhey and Wilczek\ \cite{HW}  found  that the Xanthopulos solution one gets
from the reduction of 
the polar equations of 
Schwarzschild and  Reissner-Nordstr\"{o}m black holes are \emph{pure gauge} solutions
i.e. gauge equivalent to the null perturbation. 
The  general form of the metric (\ref{metpert}) has, indeed, a gauge degree of freedom, 
{\it i.e.} there 
exists a one parameter 
class of coordinate transformations which leaves the 
form of the metric unchanged.
In the case of the GHS black hole the particular 
\emph{pure gauge} solution which reduces the system
(\ref{secforb})-(\ref{dilasos}) is\ \cite{HW}
\beq \label{count:sost1}
\ze{N}&=&-\sigma^{2} e^{-\nu}\rt + e^{4\nu}\nu_{,r}(\rt e^{-\nu})_{,r},\\
\ze{L}&=&(e^{4\nu}(\rt e^{-\nu})_{,r})_{,r}-e^{4\nu}\nu_{,r}(\rt e^{-\nu})_{,r},\\
\ze{T}&=&e^{4\nu} \rt^{-1}\rt_{,r}(\rt e^{-\nu})_{,r},\\
\ze{X}&=&\frac{\mu^{2}\enu}{2\rt},\\
\ze{B_{23}}& =& \ze{T}+\ze{L}-2\mu^{-2}\ze{X},\\
\ze{\Phi}&=&\phi_{,r} e^{4\nu}(\rt e^{-\nu})_{,r}.
\eeq
Following Holzhey and Wilczek, we now introduce a new variable $S$ 
 replacing $L$
\begin{equation}
S=B_{23}-L-X,
\end{equation}
and make  the following  substitutions:
\be\label{count:sost2}
N=\ze{N}s+n,\quad B_{23}=\ze{B}_{23}s+\frac{b}{r \rt}\quad
X=\ze{X}s+\frac{x}{\rt},\quad S=\ze{S}s,\quad
\Phi = \ze{\Phi}s+\frac{p}{\rt}.
\ee
The system is now  of  order six with respect to the new variables\, $n$, $x$, $s$, $b$ and $p$
since  $s$ does  appear through its  derivatives only.
It is easy to show from Eq. (\ref{tilo1})
that the  first derivative of $s$ can be written as a linear combination of
$b$, $x$ and $p$ as follows
\begin{equation}\label{count:sp}
s_{,r} =\frac{1}{\ze{S}}\left(\frac{\rt_{,r}}{\rt^{2} r}b - \frac{\rt_{,r}}{\rt^{2}}x-
\frac{\phi_{,r}}{\rt}p\right).
\end{equation}

\begin{figure}
$$$$
\centerline{\mbox{
\psfig{figure=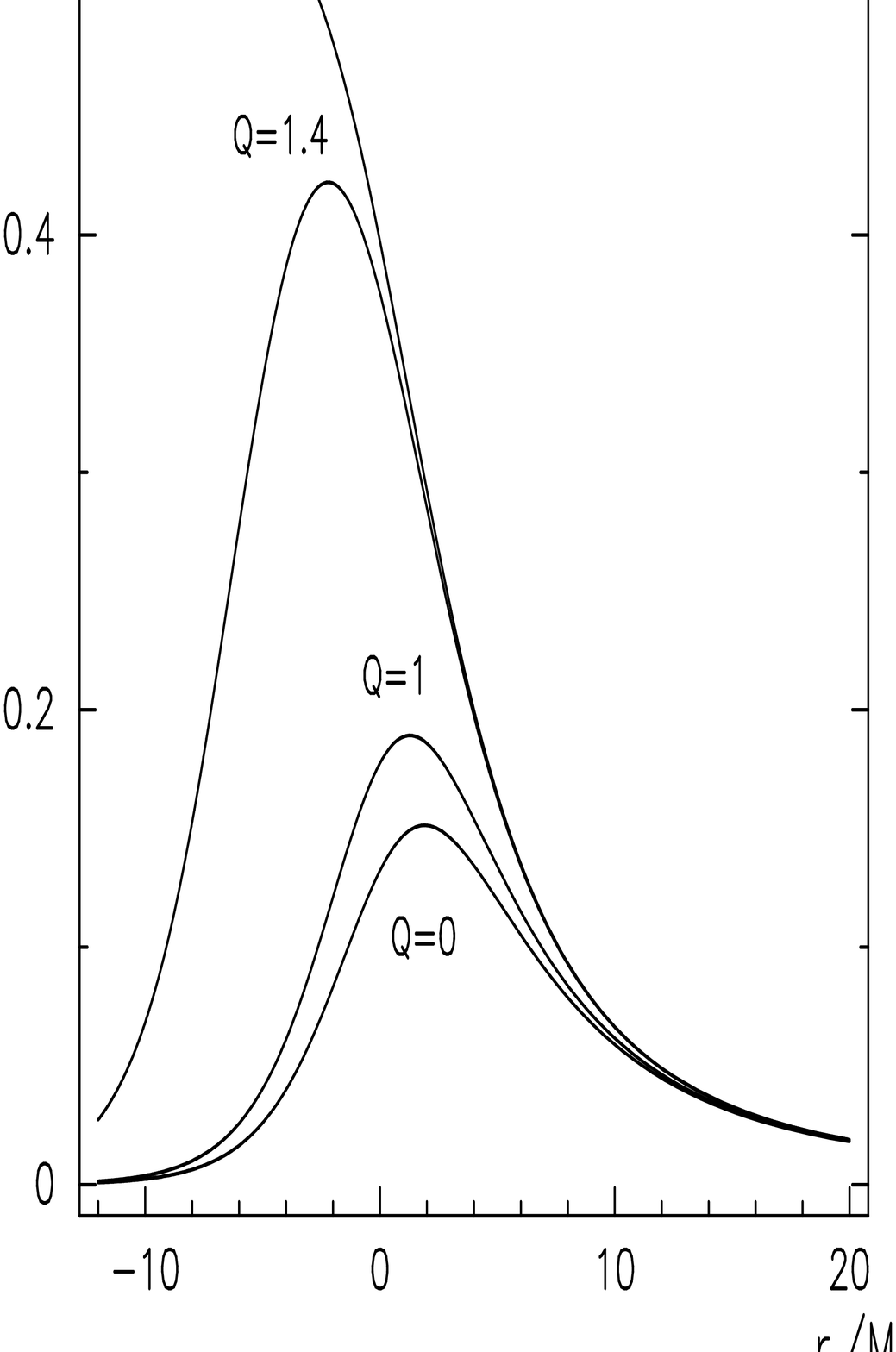,width=8cm,height=6cm}~~~~~~
\psfig{figure=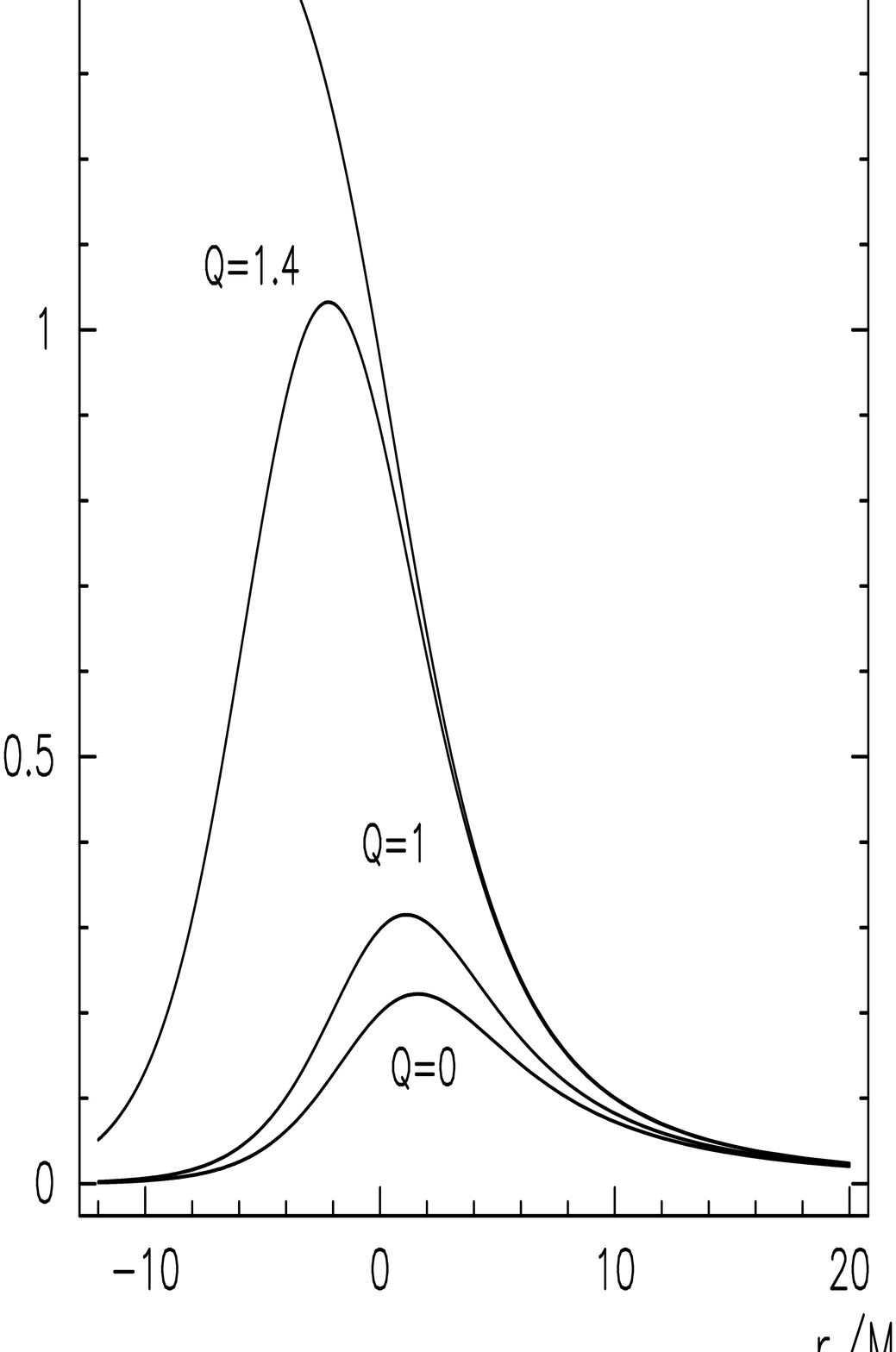,width=8cm,height=6cm}}}
\vskip 74pt
\centerline{\mbox{
\psfig{figure=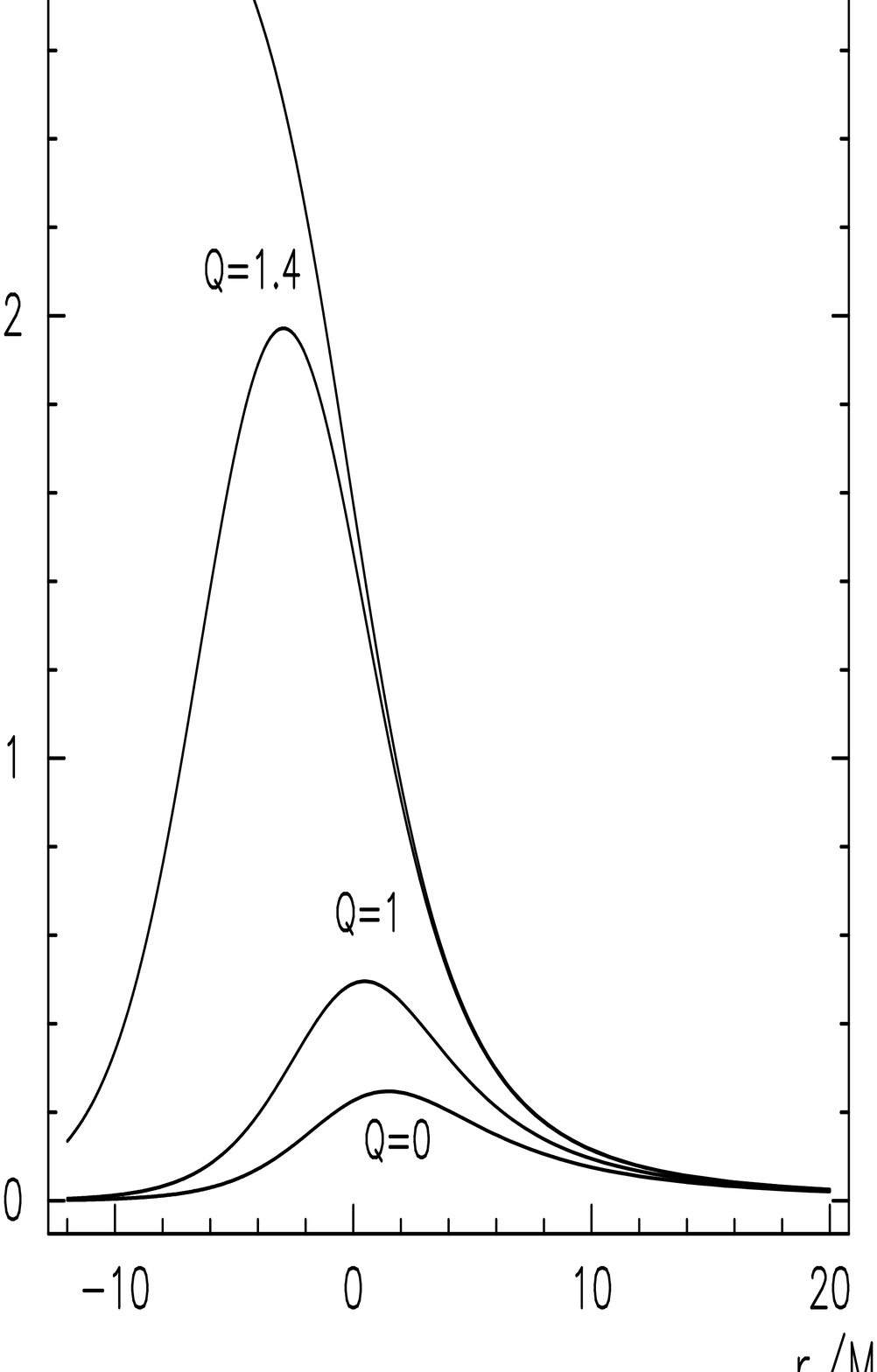,width=8cm,height=6cm}}}
\vspace{.5cm}
\caption{
The $\ell=2$ polar potentials $V_1^p$ and $V_2^p$ (upper panel)
and $V_3^p$ (lower panel)
are  plotted versus the tortoise coordinate for different values of
\ $ Q=Q_e/M.$\ When the charge is set to zero $V_1^a$ reduces to the
Zerilli potential,
whereas $V_2^p$ and $V_3^p$  are the potentials of the wave equation
describing pure electromagnetic and scalar perturbations on a
Schwarzschild background.
}
\label{fig:2}
\end{figure}

Similarly, using  (\ref{tilo2}) and  (\ref{tilo3}), we can eliminate $n$ in favour of  $x$, $b$, $p$ 
and their first derivatives. 
We are now left with only three perturbation variables,
$x$, $b$ and $p$, governed by three second order equations, namely  
(\ref{secforx}), (\ref{secforb}) and (\ref{dilasos}).
In (\ref{count:sost2}), $x$, $b$ and $p$ have been defined in such a way that their first 
derivative \emph{with respect to $r_{*}$} disappears
from the  equations they  satisfy.
Therefore the system has now the desired form
\begin{equation}\label{count:schroep}
\left(\frac{d^2}{dr_{*}^2} + \sigma^2\right){\bf v} = {\bf A} {\bf v},
\end{equation}
where 
\begin{equation}
{\bf v}=\left(\begin{array}{c}
                x\\
                b\\
                p
                \end{array}\right),
\end{equation}
and the components of the symmetric matrix {\bf A} are complicated 
functions of $r$. 

It is convenient to cast {\bf A} in the following form:
\begin{equation}\label{app:emme0}
{\bf A}(r) = \frac{1}{D(r)}\left[\, G(r) \, {\bf T}(r)\, + \, P(r)\, {\bf I}
\, \right],
\end{equation}
where the polinomials $D(r)$, $G(r)$ and $P(r)$ and the matrix ${\bf T}(r)$ are 
given in the Appendix, and ${\bf I}$ is the identity matrix. 
It is remarkable  that the eigenvectors of ${\bf T}(r)$ and  
${\bf A}(r)$ are independent of $r$.
This means that the system
can be decoupled with a basis transformation in the $x$-$b$-$p$ space. The 
three eigenvalues
are the potentials of three independent Schr\"{o}dinger equations:

\begin{equation}\label{count:eqfin2}
\left( \frac{d^{2}}{dr_{*}^{\ 2}}+\sigma^{2} \right) Z_{i}^{p}=V_{i}^{p}Z_{i}^{p} 
\hspace{10mm} (i=1,\, 2,\, 3).
\end{equation}

It should be noted  that for $\q = 0$ the equations are already decoupled 
since the off-diagonal terms
of {\bf A} vanish (see Appendix) while, as already pointed out, 
the GHS solution becomes the Schwarzschild solution.
In this case, as expected, we   find  the well-known 
potentials that rule
 pure gravitational, electromagnetic, and massless scalar perturbations
on a  Schwarzschild background, respectively.

Proving the independence of $r$ of the eigenvectors of 
${\bf T}(r)$ directly, even 
by means of a symbolic procedure, turns out to 
be awkward.
A straighforward manner is instead to expand the polinomial valued matrix as:
\begin{equation}\label{app:deco}
{\bf T} = \sum_{i=0}^{6}\, r^{i}\, {\bf T}^{(i)},
\end{equation} 
where ${\bf T}^{(i)}$ are matrices independent of $r$. It is now very easy to 
show, for instance  with the aid of MATHEMATICA,
 that the ${\bf T}^{(i)}$ commute among themselves,
so that they can be diagonalized by the same linear transformation.

The eigenvalue problem for ${\bf A}(r)$ is now considerably 
reduced by means of (\ref{app:emme0}) and (\ref{app:deco}). 
In fact, the three potentials 
can easily be obtained by fixing \emph{first} 
the values of the parameters $M$, $\q$ and $\mu$, then calculating the 
eigenvalues of each ${\bf T}^{(i)}$, which are now real numbers, and finally summing 
them up as coefficients of their respective powers of $r$. The eigenvectors of 
${\bf A}(r)$
are then obtained straighforwardly by means of (\ref{app:emme0}).
The three potentials obtained with the above procedure appear in a rather simple form
as functions of $r$ only. This fact turns out to be extremely useful, 
since to calculate the quasinormal 
frequencies by means of WKB method one needs up to their 
 sixth derivatives. 
The potentials \op V_i^p\cl associated to the wavefunctions \op Z_i^p\cl
(see Eq.  \ref{count:schroep}) are plotted in Fig.\ \ref{fig:2}
for different values of the charge.
As in the axial case, when the charge approaches the extremal value, 
all potentials assume the form of a step. 
When \op \q \rightarrow 0, \cl all off-diagonal components of the matrix
\op \bf A\cl vanish, and the system of equations 
(\ref{count:schroep})  decouple into three wave equations for the
functions $ (x, b, p)$.  From eqs.
(\ref{count:sost2}) it is easy to check  that, 
since \op B_{23}^{(0)}\cl and \op \Phi^{(0)}\cl vanish,  
the functions \op b\cl and \op \Phi\cl become purely electromagnetic and
scalar, respectively. Conversely, the variable \op x\cl is not purely
gravitational, since it contains \op B_{23}.\cl
If, however, we explicitly write the function \op x\cl in terms of \op
B_{23}, ~X\cl and \op L\cl in
the first Eq. (\ref{count:schroep}),
the electromagnetic terms  disappear  by virtue of the equation 
satisfied by $b$, and we are left with the Zerilli equation.

\section{THE QUASI-NORMAL MODES} \label{sec:5}

We have computed the complex  frequencies of the quasi-normal modes
associated to the axial
and polar wave equation by using a   WKB approximation devised by
Schutz and Will\ \cite{shutzwill} and extended to higher orders by Iyer and Will
\ \cite{iyerwill}. This approach has  been applied to the Schwarzchild\ \cite{iyer} 
and Reissner Nordstr\"{o}m\ \cite{kokkoshutz} cases, and for the 
fundamental quadrupole
mode $l=2$ agrees with other approaches\ \cite{vari} within 1\% both for the 
real and the 
immaginary parts of the first 3-4 modes. The agreement improves with 
increasing angular harmonic and decreasing mode number.

The results of our calculations are shown in table \ref{tab:1} and \ref{tab:2}, 
where we
tabulate the real and imaginary part of the
frequencies of the first five quasi normal modes associated
to the axial and polar potentials,   for different 
values of the harmonic index  $\ell$, and of the charge $\q$.

\begin{figure}
$$ $$
\centerline{\mbox{
\psfig{figure=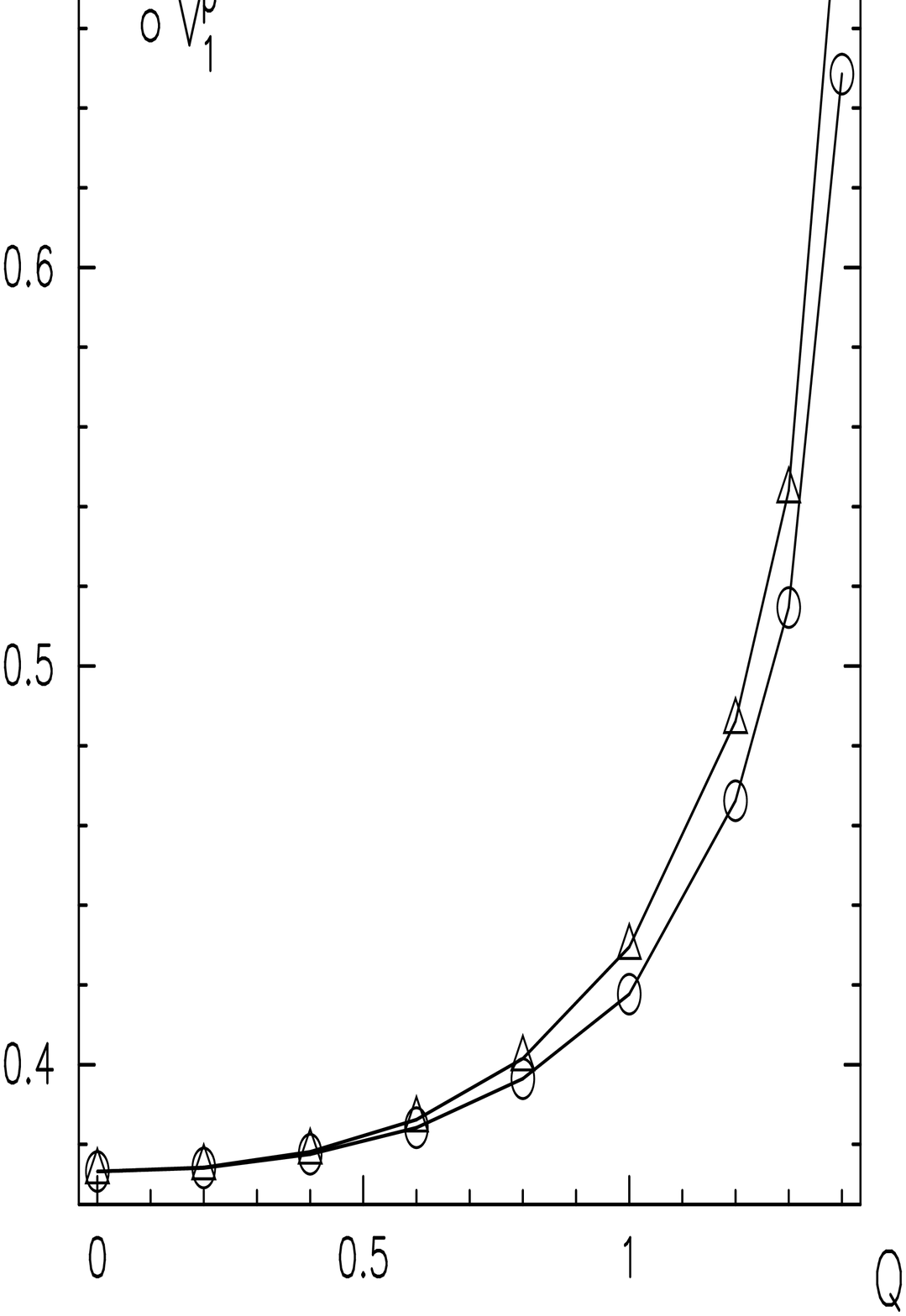,width=7.8cm,height=5.8cm}
\psfig{figure=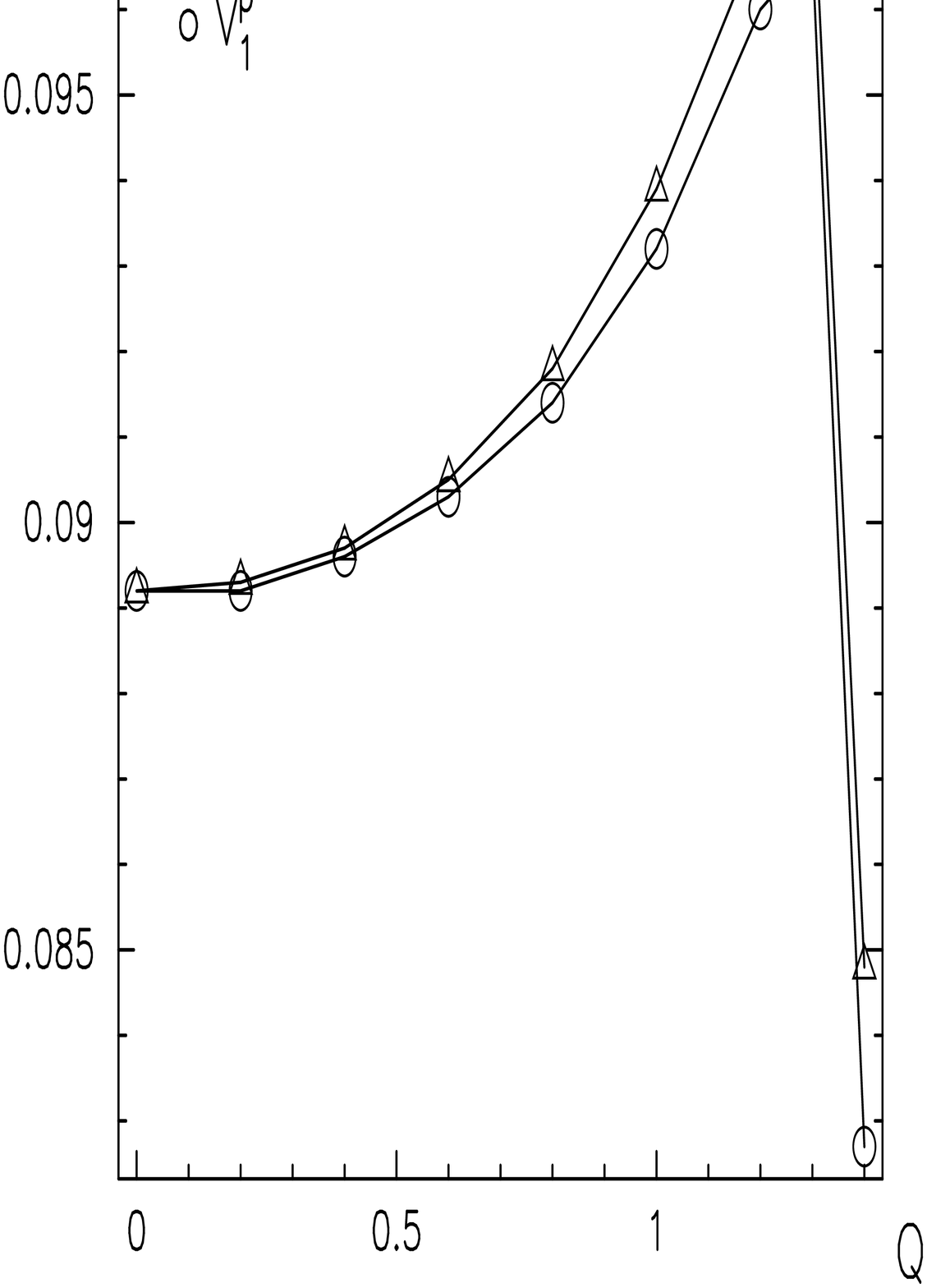,width=7.8cm,height=5.8cm}}}
\vskip 6pt
\caption{
The  real and imaginary part of the frequencies of the quasi-normal 
modes of the dilaton black holes are plotted versus the electric charge
\ $ Q=Q_e/M$, for  
$\ell=2$, and for the two potentials which reduce 
to the Regge-Wheeler and to the Zerilli potentials when the charge is set
to zero, i.e.$\ V_2^a\ $(triangles) and $\  V_1^p,$ (circles). 
}
\label{fig:3}
\end{figure}

The general behaviour of the quasi-normal frequencies for increasing values 
of the charge is well described in Figure \ref{fig:3}. There
we plot the real and the imaginary part of the first eigenfrequency of the
$\ell=2$ mode, associated to the
potentials\op V_2^a\cl and \op V_1^p,\cl as a
function of the charge $\q$.
It should be reminded that in the limit \op Q_e=0,\cl  these potentials
reduce to the Regge-Wheeler   and to the Zerilli  potentials    for
the axial and polar gravitational perturbations of a 
Schwarzschild black hole, respectively.
It is well known that the axial and polar perturbations of
a Schwarzschild black hole is a \emph{isospectral}, and this is the reason
why, in the limit \op Q_e=0,\cl  
the axial ($V_2^a$) and polar ($V_1^p$)
frequencies shown in Fig. \ref{fig:3} converge to a unique value. 

The upper value of \op Q_e/M\cl we consider in our calculations 
is \op Q_e/M= 1.4,\cl close to the limiting value 
\op Q_e/M= \sqrt{2},\cl where the potential barrier becomes a step which
reflects all incident waves.

\vspace{.8cm}

\begin{table}
\caption{The frequencies of the first five quasi normal modes associated
to the axial and polar wave equations,  
are written down in units of $1/M$, for 
$l=2$ and for increasing values of the charge $\q$.}
\begin{tabular}{dddddddddd}\tableline
$V_{1}^{a}$&&           $V_{2}^{a}$ &&$V_{1}^{p}$&&$V_{2}^{p}$&&$V_{3}^{p}$&\\ 
&&&&&&&&&\\
Re($\sigma$)&Im($\sigma$)&Re($\sigma$)&Im($\sigma$)&Re($\sigma$)
&Im($\sigma$)&Re($\sigma$)&Im($\sigma$)&Re($\sigma$)&Im($\sigma$)\\ \tableline
\multicolumn{10}{c}{\bfseries $l=2$ \ \ \ $\q=0.2$}\\ \tableline
0.374& 0.089& 0.462& 0.095&0.374&0.089&0.457&0.095&0.492&0.097\\
0.347& 0.275& 0.441& 0.292&0.346&0.275&0.436&0.291&0.472&0.297\\
0.304& 0.472& 0.408& 0.497&0.302&0.471&0.402&0.496&0.442&0.506\\
0.249& 0.674& 0.367& 0.708&0.244&0.674&0.360&0.706&0.403&0.719\\
0.180& 0.880& 0.317& 0.921&0.173&0.880&0.309&0.918&0.358&0.935\\
\tableline
\multicolumn{10}{c}{\bfseries $l=2$ \ \ \ $\q=0.4$}\\ \tableline
0.378& 0.090& 0.479& 0.096&0.377&0.090&0.461&0.095&0.516&0.099\\
0.351& 0.276& 0.458& 0.295&0.350&0.276&0.440&0.292&0.497&0.301\\
0.308& 0.473& 0.427& 0.502&0.305&0.473&0.406&0.497&0.468&0.512\\
0.253& 0.676& 0.387& 0.714&0.248&0.676&0.365&0.707&0.432&0.728\\
0.185& 0.883& 0.339& 0.929&0.177&0.883&0.315&0.920&0.389&0.945\\
\tableline
\multicolumn{10}{c}{\bfseries $l=2$ \ \ \ $\q=0.6$}\\ \tableline
0.386& 0.090& 0.508& 0.098&0.384&0.090&0.474&0.096&0.552&0.101\\
0.360& 0.279& 0.489& 0.300&0.356&0.278&0.453&0.294&0.535&0.307\\
0.318& 0.477& 0.459& 0.510&0.312&0.476&0.421&0.501&0.508&0.521\\
0.264& 0.681& 0.422& 0.725&0.255&0.680&0.381&0.712&0.476&0.740\\
0.197& 0.889& 0.378& 0.942&0.184&0.889&0.333&0.926&0.437&0.960\\
\tableline
\multicolumn{10}{c}{\bfseries $l=2$ \ \ \ $\q=0.8$}\\ \tableline
0.401& 0.092& 0.553& 0.101& 0.396& 0.091&0.500& 0.098&0.607& 0.104\\
0.376& 0.282& 0.536& 0.308& 0.369& 0.281&0.481& 0.299&0.592& 0.315\\
0.335& 0.483& 0.509& 0.522& 0.325& 0.482&0.450& 0.508&0.568& 0.533\\
0.283& 0.689& 0.476& 0.741& 0.269& 0.689&0.413& 0.722&0.539& 0.756\\
0.220& 0.898& 0.437& 0.962& 0.200& 0.899&0.368& 0.939&0.506& 0.980\\
\tableline
\multicolumn{10}{c}{\bfseries $l=2$ \ \ \ $\q=1$}\\ \tableline
0.429& 0.094& 0.627& 0.105&0.418& 0.093&0.545& 0.100&0.688&  0.107\\
0.406& 0.288& 0.612& 0.318&0.392& 0.286&0.528& 0.306&0.675&  0.324\\
0.369& 0.492& 0.589& 0.537&0.350& 0.490&0.500& 0.520&0.655&  0.547\\
0.322& 0.701& 0.561& 0.762&0.297& 0.700&0.466& 0.738&0.631&  0.775\\
0.264& 0.913& 0.529& 0.988&0.232& 0.913&0.426& 0.958&0.603&  1.004\\
\tableline
\multicolumn{10}{c}{\bfseries $l=2$ \ \ \ $\q=1.2$}\\ \tableline
0.486& 0.097& 0.762& 0.109&0.466&  0.096&0.644& 0.105&0.848& 0.111\\
0.468& 0.297& 0.752& 0.330&0.445&  0.294&0.631& 0.318&0.840& 0.336\\
0.439& 0.503& 0.736& 0.556&0.413&  0.502&0.610& 0.537&0.826& 0.565\\
0.405& 0.715& 0.717& 0.787&0.375&  0.714&0.585& 0.761&0.810& 0.799\\
0.365& 0.929& 0.696& 1.019&0.328&  0.929&0.556& 0.987&0.793& 1.034\\
\tableline
\multicolumn{10}{c}{\bfseries $l=2$ \ \ \ $\q=1.4$}\\ \tableline
0.704& 0.085& 1.252& 0.095&0.648& 0.083&1.014& 0.093&1.403& 0.096\\
0.716& 0.258& 1.259& 0.286&0.666& 0.253&1.021& 0.280&1.409& 0.288\\
0.744& 0.438& 1.274& 0.480&0.706& 0.434&1.038& 0.471&1.423& 0.483\\
0.790& 0.624& 1.297& 0.676&0.771& 0.624&1.067& 0.666&1.444& 0.679\\
0.855& 0.816& 1.331& 0.875&0.863& 0.823&1.107& 0.865&1.474& 0.879\\
\tableline
\end{tabular}\label{tab:1}
\end{table}

\begin{figure}
$$$$
$$ $$
\centerline{\mbox{
\psfig{figure=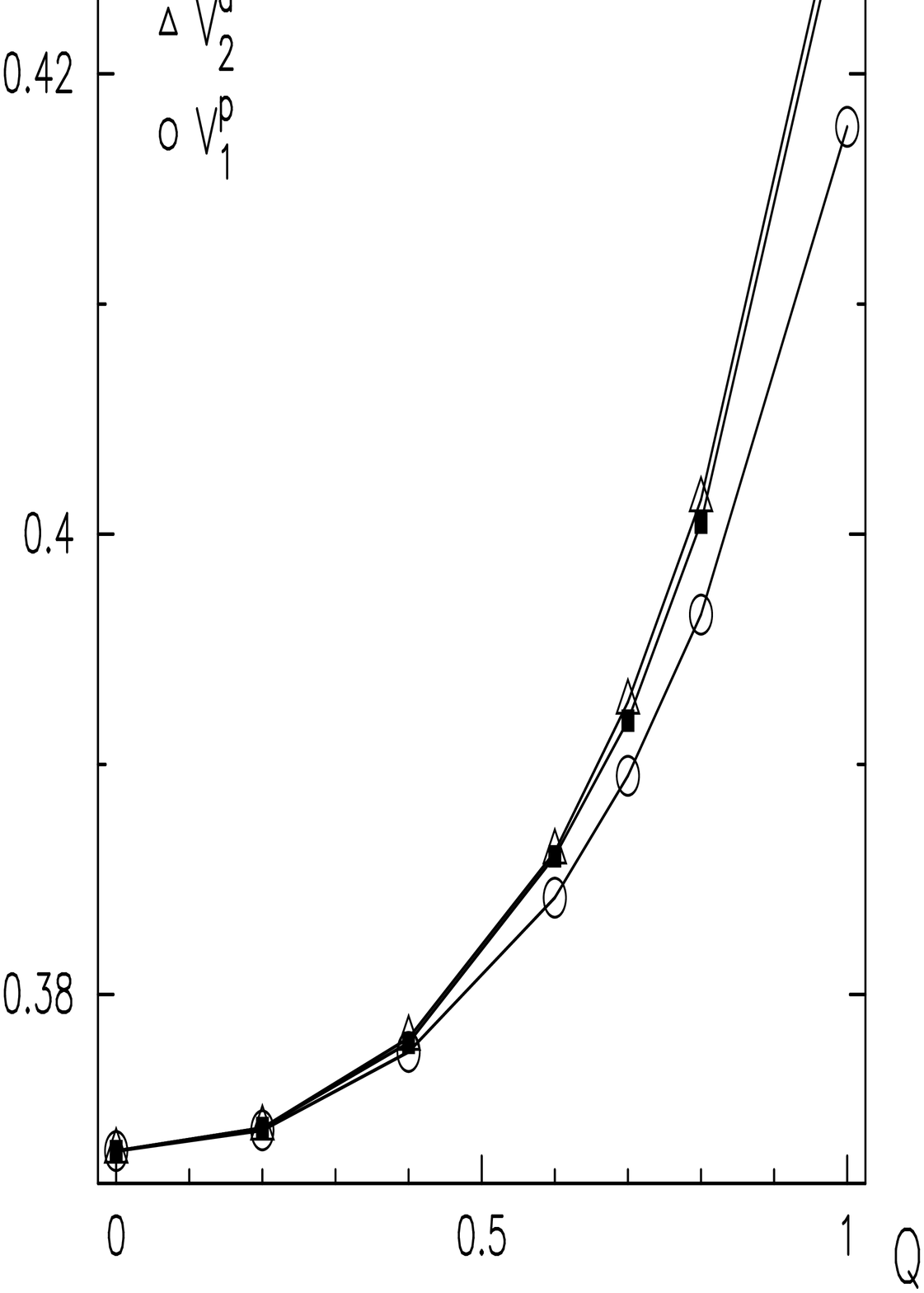,width=7.8cm,height=5.8cm}~~~~~~
\psfig{figure=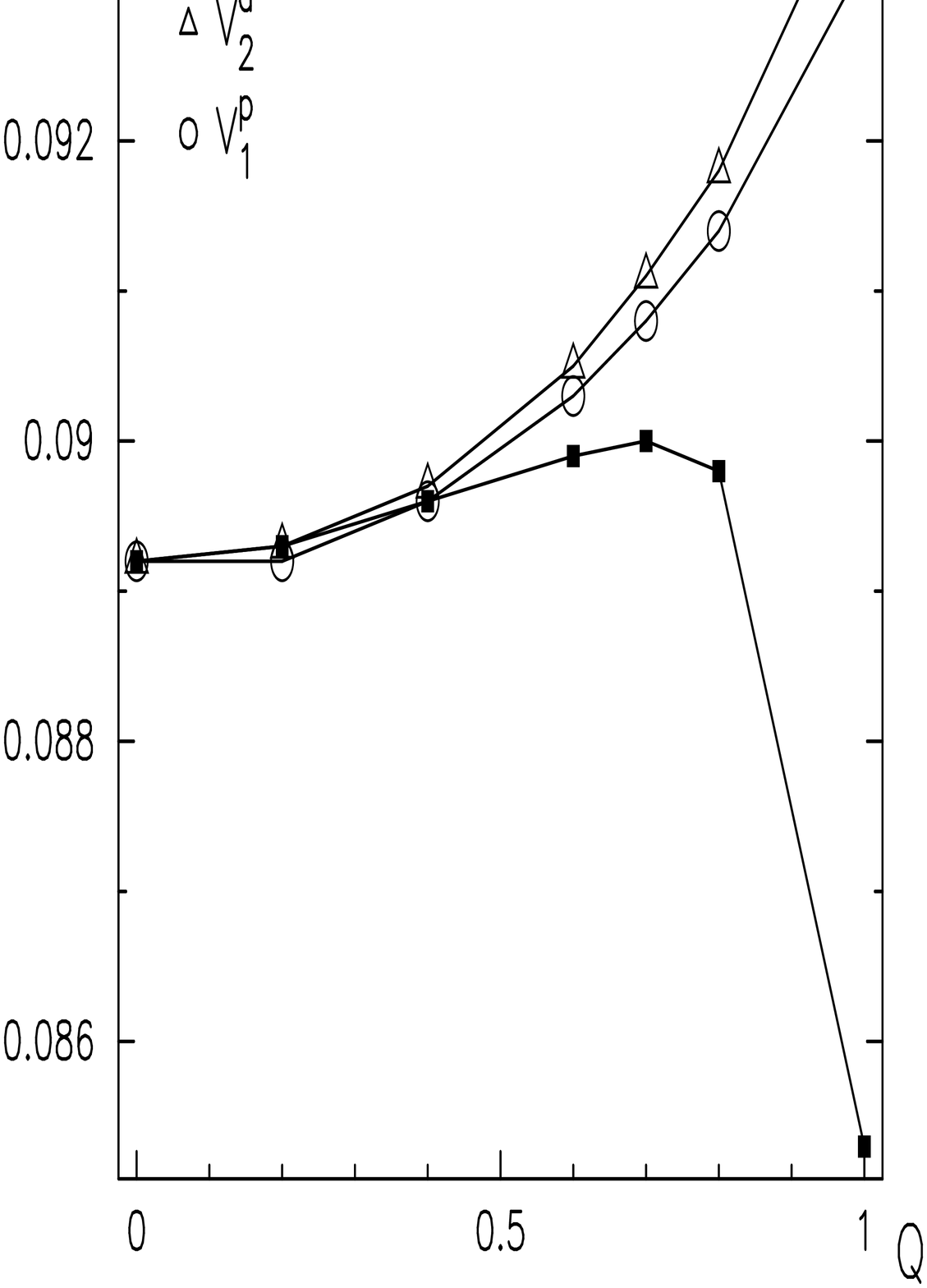,width=7.8cm,height=5.8cm}}}
\caption{
The   eigenfrequencies of the quasi-normal
modes of the dilaton black hole given in Figure 3,  are compared 
with those of a Reissner-Nordstr\"{o}m black hole for $l=2$. For $\q = 0$
both solutions reduce to Schwarzschild and the lines converge
to its first, $l=2$, pure gravitational mode $\sigma = 0.373 + 0.089 i$. For increasing 
values of the charge Reissner-Nordstr\"{o}m frequencies remain 
\emph{isospectral},
while GHS frequencies split in two parts, 
depending on whether they belong to axial 
or polar modes (see text).
Since the extremal value
of the charge for a Reissner-Nordstr\"om black hole
 is \op Q_e/M=1,\cl the data stop at that value.
}
\label{fig:4}
\end{figure}

For a Reissner-Nordstr\"om black hole the perturbed  axial and polar
equations can be decoupled in terms of four functions \op Z^{\mp}_1\cl and
\op Z^{\mp}_2$, respectively\ \cite{MT}.
In the limit \op Q_e=0,\cl the functions \op Z^{\mp}_2$  reduce to the 
Regge-Wheeler and to the Zerilli functions, whereas \op Z^{\mp}_1\cl
reduce to pure axial and polar electromagnetic functions.
Moreover, the potentials governing  \op Z^{-}_1$ and \op Z^{+}_1$  are
\emph{isospectral}, as well as those for \op Z^{-}_2$ and \op Z^{+}_2$ .
In Figure \ref{fig:4} the same data of Figure \ref{fig:3} are compared
with the lowest quasi-normal
mode  frequency of the functions $ Z^{\mp}_2$ of a Reissner-Nordstr\"om
black hole.
Since the extremal value
of the charge for a Reissner-Nordstr\"om black hole
 is \op Q_e=1,\cl the data stop at that value.

\vspace{1cm}

\begin{table}
\caption{The frequencies of the first five axial and polar quasi normal modes  
in units $1/M$ are tabulated for $l=3$ and $l=4$, for some values of the 
charge $\q$}
\begin{tabular}{dddddddddd}\tableline
$V_{1}^{a}$&&           $V_{2}^{a}$ &&$V_{1}^{p}$&&$V_{2}^{p}$&&$V_{3}^{p}$&\\ 
&&&&&&&&&\\
Re($\sigma$)&Im($\sigma$)&Re($\sigma$)&Im($\sigma$)&Re($\sigma$)&
Im($\sigma$)&Re($\sigma$)&Im($\sigma$)&Re($\sigma$)&Im($\sigma$)\\ \tableline
\multicolumn{10}{c}{\bfseries $l=3$ \ \ \ $\q=0.2$}\\ \tableline
0.601& 0.093& 0.664& 0.096&0.601&0.093&0.657&0.096&0.687&0.097\\
0.584& 0.282& 0.649& 0.291&0.584&0.282&0.642&0.290&0.672&0.294\\
0.555& 0.477& 0.623& 0.492&0.555&0.477&0.616&0.490&0.647&0.497\\
0.517& 0.678& 0.590& 0.698&0.518&0.678&0.582&0.696&0.615&0.704\\
0.473& 0.882& 0.550& 0.907&0.473&0.882&0.542&0.904&0.578&0.915\\
\tableline
\multicolumn{10}{c}{\bfseries $l=3$ \ \ \ $\q=.4$}\\ \tableline
0.607& 0.093& 0.686& 0.097&0.606&0.093&0.666&0.096&0.716&0.098\\
0.590& 0.283& 0.672& 0.294&0.589&0.282&0.651&0.291&0.702&0.298\\
0.562& 0.479& 0.647& 0.496&0.560&0.478&0.625&0.492&0.678&0.503\\
0.525& 0.680& 0.614& 0.704&0.523&0.680&0.592&0.698&0.648&0.712\\
0.480& 0.885& 0.577& 0.915&0.479&0.885&0.552&0.908&0.612&0.926\\
\tableline
\multicolumn{10}{c}{\bfseries $l=3$ \ \ \ $\q=.6$}\\ \tableline
0.621& 0.094& 0.724& 0.099&0.617&0.094&0.688&0.097&0.762&0.100\\
0.605& 0.285& 0.711& 0.299&0.601&0.284&0.673&0.294&0.749&0.303\\
0.576& 0.482& 0.687& 0.504&0.572&0.481&0.648&0.497&0.726&0.511\\
0.540& 0.685& 0.657& 0.715&0.536&0.684&0.616&0.705&0.698&0.724\\
0.497& 0.891& 0.622& 0.928&0.492&0.889&0.579&0.916&0.665&0.940\\
\tableline
\multicolumn{10}{c}{\bfseries $l=4$ \ \ \ $\q=0.2$}\\ \tableline
0.811& 0.094& 0.862& 0.096&0.811&0.094&0.854&0.096&0.882&0.097\\
0.799& 0.285& 0.851& 0.290&0.798&0.285&0.842&0.289&0.870&0.292\\
0.776& 0.479& 0.801& 0.691&0.776&0.479&0.821&0.487&0.849&0.492\\
0.746& 0.679& 0.768& 0.898&0.745&0.679&0.792&0.690&0.822&0.696\\
0.710& 0.882& 0.729& 1.107&0.709&0.882&0.758&0.895&0.789&0.904\\
\tableline
\multicolumn{10}{c}{\bfseries $l=4$ \ \ \ $\q=0.4$}\\ \tableline
0.820& 0.095& 0.890& 0.097&0.818&0.094&0.868&0.096&0.916&0.098\\
0.808& 0.286& 0.879& 0.293&0.806&0.285&0.856&0.291&0.905&0.296\\
0.785& 0.481& 0.858& 0.493&0.783&0.481&0.835&0.490&0.885&0.498\\
0.755& 0.681& 0.831& 0.698&0.753&0.681&0.807&0.693&0.858&0.704\\
0.720& 0.885& 0.798& 0.906&0.718&0.884&0.774&0.899&0.827&0.913\\
\tableline
\multicolumn{10}{c}{\bfseries $l=4$ \ \ \ $\q=0.6$}\\ \tableline
0.841& 0.095& 0.936& 0.099&0.835&0.095&0.899&0.097&0.969&0.100\\
0.829& 0.288& 0.926& 0.298&0.823&0.287&0.888&0.294&0.959&0.301\\
0.807& 0.485& 0.906& 0.501&0.801&0.484&0.867&0.495&0.941&0.506\\
0.778& 0.686& 0.881& 0.708&0.771&0.685&0.840&0.700&0.916&0.715\\
0.743& 0.891& 0.850& 0.919&0.736&0.890&0.808&0.908&0.887&0.927\\
\tableline
\end{tabular}\label{tab:2}
\end{table}

We see that the  real part of the frequency  increases as a 
function of the charge
both for a dilaton, and for a Reissner-Nordstr\"om black hole; similarly,
the imaginary part increases and then decreases to a finite value as the
charge approaches the limiting value. Since however 
since the limiting values are different, 
the imaginary part for Reissner-Nordstr\"om begins to decrease while the 
corresponding GHS values still increase.

Figures \ref{fig:3} and \ref{fig:4} clearly show that
 the potentials \op V_2^a\cl and
\op V_1^p\cl  of a dilaton black hole \emph{are not isospectral} as they are
instead when  \op Q_e=0.\cl 
In addition, it should be reminded that the potential \op V_2^a\cl
rules the equation for \op Z_2^a\cl which is a combination of
electromagnetic and gravitational perturbations only 
(cfr. Eq. \ref{transf}), whereas
\op V_1^p\cl appears in the polar equation for  \op Z_1^p,\cl
which is a combination of electromagnetic, gravitational and scalar
perturbations.

\begin{figure}
$$ $$
$$ $$
\centerline{\mbox{
\psfig{figure=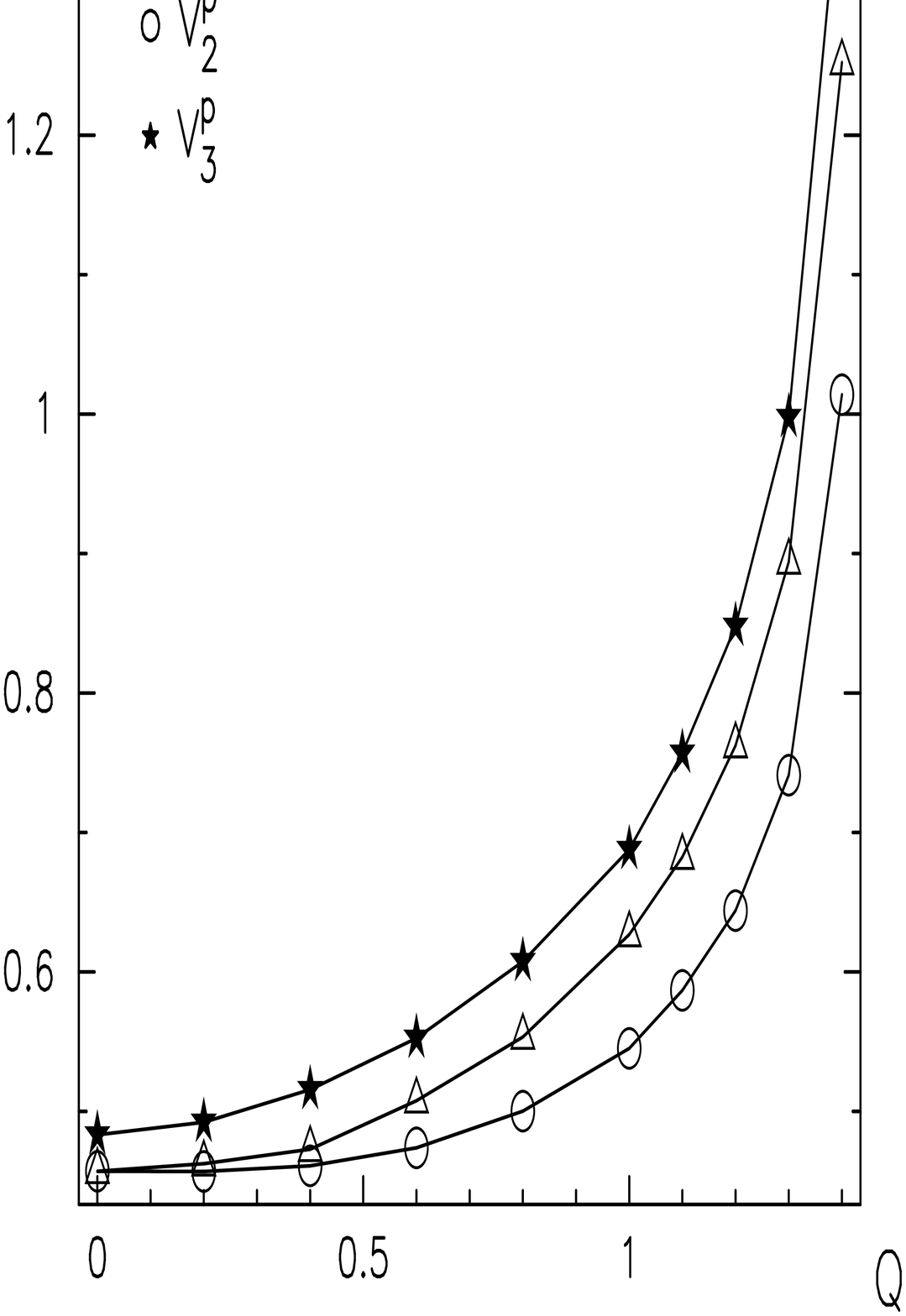,width=7.8cm,height=5.8cm}~~~~~~
\psfig{figure=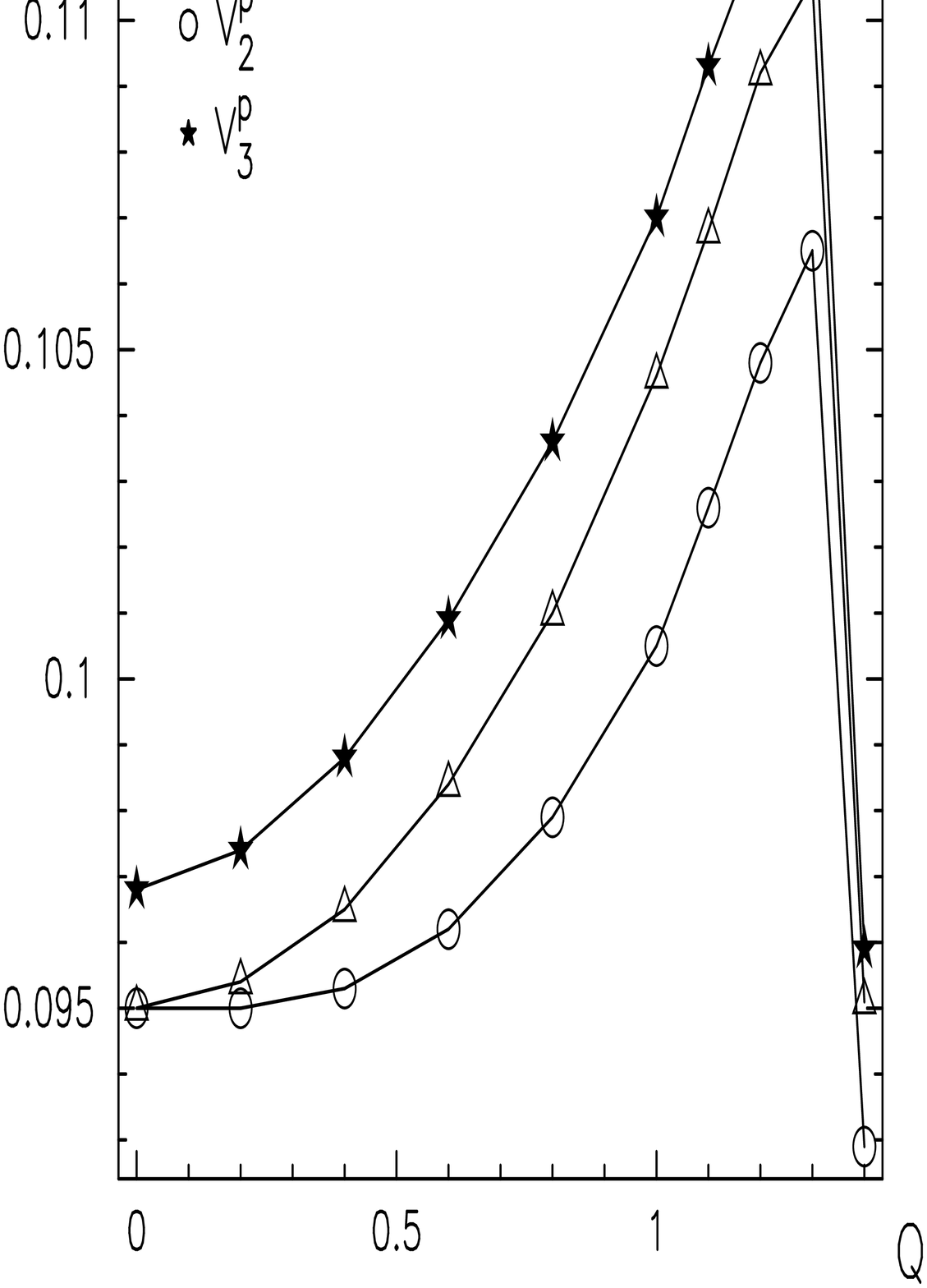,width=7.8cm,height=5.8cm}}}
\vskip 14pt
\caption{
The  real and imaginary part of the frequencies of the lowest 
$\ell=2$-quasi-normal modes of the dilaton black hole, 
are plotted versus the electric charge \ $ Q=Q_e/M$, 
for the three potentials 
\  $ V_1^a$ (triangles), $\ V_2^p$ (circles)\  and \ $\ V_3^p$ (stars).
In the limit \ $ Q=0$, 
$V_1^a$ and $\ V_2^p$ reduce  to the potentials governing
the axial and   polar pure electromagnetic perturbations
of a Schwarzschild black hole, whereas $\ V_3^p$ becomes the potential
which appears in the wave equation of  the  pure scalar perturbations. 
}
\label{fig:5}
\end{figure}

Finally, in Figure \ref{fig:5} we plot the real and imaginary part 
of the frequency of the lowest $l=2$ quasi-normal mode for the remaining
potentials, \op V_1^a,\cl \op V_2^p\cl and \op V_3^p,\cl as functions of
the charge \op Q_e.\cl
When \op Q_e=0,\cl $V_1^a$ and  $V_2^p$ reduce to those governing the pure
electomagnetic, axial and polar perturbations of a Schwarzschild black hole,
respectively, and indeed, they are isospectral in that limit.
In the same limit, \op V_3^p\cl reduces to the potential of the pure scalar
perturbations of a Schwarzschild black hole.

\section{CONCLUDING REMARKS}\label{sec:6}

The spectrum of the quasi-normal modes of  a charged, dilaton black hole 
is different from that of a Schwarzschild or a Reissner-Nordstr\"om black
hole.
For a Schwarzschild black hole the perturbations are completely described
by the Regge-Wheeler and the Zerilli equations for  the axial
and the polar perturbations. Although the analytic form of the
two potential barriers is different, they are related by a very simple
equation (MT, ch. 5, \& 43) which allows them to have the same reflexion and
trasmission coefficients. Therefore, since the quasi-normal mode frequencies 
are the singularities of the scattering amplitude, it follows that the two
potentials are \emph{isospectral}. 
Thus, a  perturbed Scharzschild black hole
emits axial and polar gravitational waves at exactly 
the same frequencies.

For a Reissner-Nordstr\"om black hole the perturbed equations
can be reduced to  four wave equations, two for the 
axial and two for the polar perturbations, respectively. The four
wave-functions \op Z^\pm_1\cl and \op
Z^\pm_2,\cl where $+$ stands for polar and $-$ for axial,
are a linear combination of the gravitational and
the electromagnetic functions belonging to the corresponding parity.
It turns out that the  two potentials\op V^+_1\cl and\op V^-_1\cl
are again related in such a way that they have the same reflection and
trasmission coefficients, and the same is true for 
\op V^\pm_2.\cl 
Thus the coupling GW-EM is such that it preseves the isospectrality  of the
axial and polar perturbations.  However there is an important difference
with respect to the Schwarzschild case: 
no quasi-normal mode exists that is purely
electromagnetic or gravitational, which means that the excitation of a
mode will be accompanied by the simultaneous emission of
both gravitational and electromagnetic waves.

For a charged black hole in a theory described by action (\ref{theaction}) the 
situation is
different. Let us consider the axial perturbations first.
As shown in Section 3, the perturbed equations can be reduced to two wave
equations, but the  perturbed dilaton does not couple
to the electromagnetic and gravitational fields.
This is due to the fact that the dilaton is a scalar, and its axial
perturbation vanishes.
Consequently, the excitation of an axial mode will be
accompanied only by the emission of gravitational and electromagnetic
waves. However, the dilaton appears in the unperturbed metric functions
that determine the shape of the potentials of the axial wave equations.
Thus it  affects the scattering properties of the axial potentials.
It is interesting to note that the real part of the quasi-normal mode
frequencies of  the axial potential \op V_2^a\cl are very similar 
to those of the Reissner-Nordstr\"om black hole (see Fig. \ref{fig:4}), even 
though the dilaton solution does not 
reduce to the Reisnner-Nordstr\"om solution in any limiting case.

On the other hand, the two wave equations which describe the polar perturbations of a 
Reissner-Nordstr\"om black hole, are replaced by three  wave equations
in the case of a dilaton black hole, and they couple the gravitational,
electromagnetic and scalar perturbations. This occurrence breaks the
symmetry between axial and polar perturbations, and makes the scattering
properties of the two parities different (see Fig. \ref{fig:5}).

We conclude that for a dilaton black hole
the excitation of an axial mode induces the simultaneous emission of
gravitational and electromagnetic waves, whereas the excitation of a polar
mode is accompanied by  the further emission of scalar radiation.
In addition, gravitational and  electromagnetic polar waves
are emitted with   frequencies and damping times
different from the axial ones.

\section*{Acknowledgments}

F.P. whishes to thank Roberto De Pietri, Antonio Scotti and Michele Vallisneri for 
useful discussions and help in computing technicalities.

\section{APPENDIX} 

In this Appendix we provide all the elements of the matrix {\bf A} 
that appears in equation 
(\ref{count:schroep}) and in which are contained all the relevant 
features of the polar perturbations. 
It proves convenient to cast {\bf A} in the following form
 \begin{equation}\label{app:emme}
{\bf A}(r) =\frac{1}{D(r)}\left[\, G(r) \, {\bf T}(r)\, + \, P(r)\, {\bf I}
\, \right],
\end{equation}
where {\bf I} is the identity matrix, and $D(r)$, $G(r)$ and ${\bf T}(r)$ are:
 \begin{eqnarray*}
D(r)& =&
4\,\mu \,\q\,{r^4}\,{{\left( -{{\q}^2} + M\,r \right) }^2}\\
&&  \times\left( 3\,M\,{{\q}^4} - 7\,{M^2}\,{{\q}^2}\,r - {{\q}^4}\,r + 6\,{M^3}\,{r^2} - 
    M\,{{\mu }^2}\,{{\q}^2}\,{r^2} + {M^2}\,{{\mu }^2}\,{r^3} \right) \\
&& \times{{\left( 4\,M\,{{\q}^4} - 14\,{M^2}\,{{\q}^2}\,r - {{\q}^4}\,r + 12\,{M^3}\,{r^2} - 
       2\,M\,{{\mu }^2}\,{{\q}^2}\,{r^2} + 2\,{M^2}\,{{\mu }^2}\,{r^3} \right) }^2},\\[3mm]
 G(r)& =& \mu {\q}\left( 2\,M - r \right) \,{r^2}
 \times \left( -{{\q}^2} + M\,r \right) \\ 
 && \left( 3\,M\,{{\q}^4} - 7\,{M^2}\,{{\q}^2}\,r - {{\q}^4}\,r + 
     6\,{M^3}\,{r^2} - M\,{{\mu }^2}\,{{\q}^2}\,{r^2} + {M^2}\,{{\mu }^2}\,{r^3} \right)/M,\\[3mm] 
P(r)&=&(1936\,{M^6}\,{{\q}^6} + 324\,{M^4}\,{{\q}^8} -  
     160\,{M^4}\,{{\mu }^2}\,{{\q}^8} + 8\,{M^2}\,{{\q}^{10}} + {{\q}^{12}} -  \\
& &    2432\,{M^7}\,{{\q}^4}\,r - 448\,{M^5}\,{{\q}^6}\,r + 
      624\,{M^5}\,{{\mu }^2}\,{{\q}^6}\,r + 88\,{M^3}\,{{\mu }^2}\,{{\q}^8}\,r + \\
& &     2112\,{M^8}\,{{\q}^2}\,{r^2} - 1136\,{M^6}\,{{\q}^4}\,{r^2} - 
      1520\,{M^6}\,{{\mu }^2}\,{{\q}^4}\,{r^2} - 64\,{M^4}\,{{\q}^6}\,{r^2} - \\
& &     232\,{M^4}\,{{\mu }^2}\,{{\q}^6}\,{r^2} +  
     24\,{M^4}\,{{\mu }^4}\,{{\q}^6}\,{r^2} - 8\,{M^2}\,{{\mu }^2}\,{{\q}^8}\,{r^2} - \\
& &     1152\,{M^9}\,{r^3} + 2688\,{M^7}\,{{\q}^2}\,{r^3} + 
      1984\,{M^7}\,{{\mu }^2}\,{{\q}^2}\,{r^3} - \\
& &    272\,{M^5}\,{{\mu }^2}\,{{\q}^4}\,{r^3} - 
      264\,{M^5}\,{{\mu }^4}\,{{\q}^4}\,{r^3} + 
     16\,{M^3}\,{{\mu }^2}\,{{\q}^6}\,{r^3}\\ 
&& - 4\,{M^3}\,{{\mu }^4}\,{{\q}^6}\,{r^3} - 
      1152\,{M^8}\,{r^4} - 960\,{M^8}\,{{\mu }^2}\,{r^4} + 
      832\,{M^6}\,{{\mu }^2}\,{{\q}^2}\,{r^4} +  \\
& &    464\,{M^6}\,{{\mu }^4}\,{{\q}^2}\,{r^4} - 
      16\,{M^4}\,{{\mu }^2}\,{{\q}^4}\,{r^4} - 
      28\,{M^4}\,{{\mu }^4}\,{{\q}^4}\,{r^4} - \\
& &     16\,{M^4}\,{{\mu }^6}\,{{\q}^4}\,{r^4} - 384\,{M^7}\,{{\mu }^2}\,{r^5} - 
      224\,{M^7}\,{{\mu }^4}\,{r^5} + 64\,{M^5}\,{{\mu }^4}\,{{\q}^2}\,{r^5} + \\
& &     32\,{M^5}\,{{\mu }^6}\,{{\q}^2}\,{r^5} - 32\,{M^6}\,{{\mu }^4}\,{r^6} - 
      16\,{M^6}\,{{\mu }^6}\,{r^6})/{8\,{M^2}\,\mu \,\q}.\\[3mm]
{\bf T}_{11}&=& 
2448\,{M^6}\,{{\q}^6} + 324\,{M^4}\,{{\q}^8} - 
  160\,{M^4}\,{{\mu }^2}\,{{\q}^8} + 
  8\,{M^2}\,{{\q}^{10}} + {{\q}^{12}} + 
  8\,{M^3}\,{{\q}^4}\hspace{5cm}\\
& &  ( -496\,{M^4} - 64\,{M^2}\,{{\q}^2} + 
     110\,{M^2}\,{{\mu }^2}\,{{\q}^2} - 
     4\,{{\q}^4} + 11\,{{\mu }^2}\,{{\q}^4}
      ) \,r + 8\,{M^2}\,{{\q}^2}\\
&  & ( 408\,{M^6} + 42\,{M^4}\,{{\q}^2} - 
     222\,{M^4}\,{{\mu }^2}\,{{\q}^2} -      45\,{M^2}\,{{\mu }^2}\,{{\q}^4} + 
     3\,{M^2}\,{{\mu }^4}\,{{\q}^4} - \\
&   &  {{\mu }^2}\,{{\q}^6} ) \,{r^2} + 
  4\,{M^3}\,( -288\,{M^6} +      400\,{M^4}\,{{\mu }^2}\,{{\q}^2} + 
     108\,{M^2}\,{{\mu }^2}\,{{\q}^4} - \\
&    & 34\,{M^2}\,{{\mu }^4}\,{{\q}^4} + 
     4\,{{\mu }^2}\,{{\q}^6} -      {{\mu }^4}\,{{\q}^6} ) \,{r^3} + 
  4\,{M^4}\,{{\mu }^2}\,( -144\,{M^4} - \\
&    & 32\,{M^2}\,{{\q}^2} + 
     52\,{M^2}\,{{\mu }^2}\,{{\q}^2} -     4\,{{\q}^4} - 7\,{{\mu }^2}\,{{\q}^4} - 
     4\,{{\mu }^4}\,{{\q}^4} ) \,{r^4} + \\
& & 32\,{M^5}\,{{\mu }^4}\,
   ( -3\,{M^2} + 2\,{{\q}^2} +     {{\mu }^2}\,{{\q}^2} ) \,{r^5} - 
  16\,{M^6}\,{{\mu }^4}\,( 2 + {{\mu }^2} ) \,{r^6}\\[3mm]
{\bf T}_{21}& =& 
128\,{M^5}\,\mu \,{{\q}^7} + 
  8\,{M^2}\,\mu \,{{\q}^5}\,
   ( -76\,{M^4} + 8\,{M^2}\,{{\mu }^2}\,{{\q}^2} - \hspace{5cm}\\
&&     {{\q}^4} ) \,r + 
  16\,{M^3}\,\mu \,{{\q}^3}\,
   ( 56\,{M^4} + 22\,{M^2}\,{{\q}^2} - 
     8\,{M^2}\,{{\mu }^2}\,{{\q}^2} + {{\q}^4} - \\
&&     2\,{{\mu }^2}\,{{\q}^4} ) \,{r^2} + 
  32\,{M^4}\,\mu \,\q\,
   ( -12\,{M^4} - 24\,{M^2}\,{{\q}^2} + \\
&&     2\,{M^2}\,{{\mu }^2}\,{{\q}^2} + 
     6\,{{\mu }^2}\,{{\q}^4} + {{\mu }^4}\,{{\q}^4}
      ) \,{r^3} + 32\,{M^5}\,\mu \,\q\\
& &  ( 12\,{M^2} - 9\,{{\mu }^2}\,{{\q}^2} - 
     2\,{{\mu }^4}\,{{\q}^2} ) \,{r^4} + 
  32\,{M^6}\,{{\mu }^3}\,( 4 + {{\mu }^2} )\q\,{r^5}\\[3mm]
{\bf T}_{31}& =&
16\,{M^3}\,\mu \,{\sqrt{2 + {{\mu }^2}}}\,{{\q}^2}\,r\,
  ( {{\q}^2} - M\,r ) \,
  ( 14\,{M^2}\,{{\q}^2} - {{\q}^4} - \hspace{5cm}\\
& &   24\,{M^3}\,r + 4\,M\,{{\mu }^2}\,{{\q}^2}\,r + 
    6\,{M^2}\,{r^2} - 6\,{M^2}\,{{\mu }^2}\,{r^2} - 
    {{\mu }^2}\,{{\q}^2}\,{r^2} + 2\,M\,{{\mu }^2}\,{r^3})\\[3mm]
{\bf T}_{22} &=&
2064\,{M^6}\,{{\q}^6} + 260\,{M^4}\,{{\q}^8} - 
  160\,{M^4}\,{{\mu }^2}\,{{\q}^8} + 
  8\,{M^2}\,{{\q}^{10}} + {{\q}^{12}} + \hspace{5cm}\\
& & 4\,M\,{{\q}^4}\,( -536\,{M^6} - 
     164\,{M^4}\,{{\q}^2} + 
     116\,{M^4}\,{{\mu }^2}\,{{\q}^2} + \\
&  &   6\,{M^2}\,{{\q}^4} + 
     18\,{M^2}\,{{\mu }^2}\,{{\q}^4} + {{\q}^6}
      ) \,r + 8\,{M^2}\,{{\q}^2}\\
& &  ( 72\,{M^6} + 6\,{M^4}\,{{\q}^2} - 
     98\,{M^4}\,{{\mu }^2}\,{{\q}^2} - 
     32\,{M^2}\,{{\q}^4} - \\
& &    43\,{M^2}\,{{\mu }^2}\,{{\q}^4} - 
     5\,{M^2}\,{{\mu }^4}\,{{\q}^4} - {{\q}^6} + 
     {{\mu }^2}\,{{\q}^6} ) \,{r^2} +\\ 
& & 4\,{M^3}\,{{\q}^2}\,
   ( 432\,{M^4} + 256\,{M^4}\,{{\mu }^2} + 
     48\,{M^2}\,{{\q}^2} + \\
& &    12\,{M^2}\,{{\mu }^2}\,{{\q}^2} - 
     18\,{M^2}\,{{\mu }^4}\,{{\q}^2} - 
     12\,{{\mu }^2}\,{{\q}^4} -\\ 
& &    {{\mu }^4}\,{{\q}^4} ) \,{r^3} + 
  4\,{M^4}\,( -288\,{M^4} - 144\,{M^4}\,{{\mu }^2} + 
     160\,{M^2}\,{{\mu }^2}\,{{\q}^2} + \\
& &    76\,{M^2}\,{{\mu }^4}\,{{\q}^2} + 
     8\,{{\mu }^2}\,{{\q}^4} - 
     11\,{{\mu }^4}\,{{\q}^4} - \\
&  &   4\,{{\mu }^6}\,{{\q}^4} ) \,{r^4} + 
  16\,{M^5}\,{{\mu }^2}\,( -24\,{M^2} - 
     12\,{M^2}\,{{\mu }^2} + 5\,{{\mu }^2}\,{{\q}^2} + \\
&   &  2\,{{\mu }^4}\,{{\q}^2} ) \,{r^5} - 
  16\,{M^6}\,{{\mu }^4}\,( 2 + {{\mu }^2} ) \,{r^6}\\[3mm]
{\bf T}_{32}& =&
 8\,{M^2}\,{\sqrt{2 + {{\mu }^2}}}\,\q\,
  ( -16\,{M^3}\,{{\q}^6} - 
    36\,{M^4}\,{{\q}^4}\,r + \\
& &   8\,{M^2}\,{{\q}^6}\,r - 
    8\,{M^2}\,{{\mu }^2}\,{{\q}^6}\,r + 
    {{\q}^8}\,r + 192\,{M^5}\,{{\q}^2}\,{r^2} -\hspace{5cm}\\ 
& &   52\,{M^3}\,{{\q}^4}\,{r^2} - 
    16\,{M^3}\,{{\mu }^2}\,{{\q}^4}\,{r^2} - 
    2\,M\,{{\q}^6}\,{r^2} + \\
& &   4\,M\,{{\mu }^2}\,{{\q}^6}\,{r^2} - 
   144\,{M^6}\,{r^3} + 48\,{M^4}\,{{\q}^2}\,{r^3} + 
    72\,{M^4}\,{{\mu }^2}\,{{\q}^2}\,{r^3} - \\
& &   16\,{M^2}\,{{\mu }^2}\,{{\q}^4}\,{r^3} - 
    4\,{M^2}\,{{\mu }^4}\,{{\q}^4}\,{r^3} - 
    48\,{M^5}\,{{\mu }^2}\,{r^4} + \\
& &   12\,{M^3}\,{{\mu }^2}\,{{\q}^2}\,{r^4} + 
    8\,{M^3}\,{{\mu }^4}\,{{\q}^2}\,{r^4} - 
    4\,{M^4}\,{{\mu }^4}\,{r^5} )\\[3mm]
{\bf T}_{33} &=&
1936\,{M^6}\,{{\q}^6} + 324\,{M^4}\,{{\q}^8} - 
  160\,{M^4}\,{{\mu }^2}\,{{\q}^8} + 
  8\,{M^2}\,{{\q}^{10}} + {{\q}^{12}} + \\
& & 8\,{M^3}\,{{\q}^4}\,
   ( -304\,{M^4} - 56\,{M^2}\,{{\q}^2} + 
     78\,{M^2}\,{{\mu }^2}\,{{\q}^2} + \\
& &    11\,{{\mu }^2}\,{{\q}^4} ) \,r + 
  8\,{M^2}\,{{\q}^2}\,
   ( 264\,{M^6} - 142\,{M^4}\,{{\q}^2} - \hspace{5cm}\\
& &    190\,{M^4}\,{{\mu }^2}\,{{\q}^2} - 
     8\,{M^2}\,{{\q}^4} - 
     29\,{M^2}\,{{\mu }^2}\,{{\q}^4} + \\
& &    3\,{M^2}\,{{\mu }^4}\,{{\q}^4} - 
     {{\mu }^2}\,{{\q}^6} ) \,{r^2} + 
  4\,{M^3}\,( -288\,{M^6} + 672\,{M^4}\,{{\q}^2} + \\
& &    496\,{M^4}\,{{\mu }^2}\,{{\q}^2} - 
     68\,{M^2}\,{{\mu }^2}\,{{\q}^4} - 
     66\,{M^2}\,{{\mu }^4}\,{{\q}^4} + \\
& &    4\,{{\mu }^2}\,{{\q}^6} - {{\mu }^4}\,{{\q}^6}
      ) \,{r^3} + 4\,{M^4}\,
   ( -288\,{M^4} - 240\,{M^4}\,{{\mu }^2} + \\
& &    208\,{M^2}\,{{\mu }^2}\,{{\q}^2} + 
     116\,{M^2}\,{{\mu }^4}\,{{\q}^2} - 
     4\,{{\mu }^2}\,{{\q}^4} - \\
& &    7\,{{\mu }^4}\,{{\q}^4} - 
     4\,{{\mu }^6}\,{{\q}^4} ) \,{r^4} + 
  32\,{M^5}\,{{\mu }^2}\,( -12\,{M^2} - \\
& &    7\,{M^2}\,{{\mu }^2} + 2\,{{\mu }^2}\,{{\q}^2} + 
     {{\mu }^4}\,{{\q}^2} ) \,{r^5} - 
  16\,{M^6}\,{{\mu }^4}\,( 2 + {{\mu }^2} ) \,{r^6}
\end{eqnarray*}

\end{document}